\documentclass[aps,prx,article,twocolumn,preprintnumbers,amsmath,amssymb,superscriptaddress,longbibliography]{revtex4-1}

\date{\today}
\usepackage{epsfig}
\usepackage{subfigure}
\usepackage{graphicx}
\usepackage{dcolumn}
\usepackage{bm}
\usepackage[colorlinks,linkcolor=blue,hyperindex,CJKbookmarks]{hyperref}
\usepackage{float}
\usepackage{hyperref}
\usepackage{comment}
\usepackage{soul}
\newcommand{\real}      {\mathrm{Re}}

\newcommand{\Tr}    {{\mathrm{Tr}}}
\newcommand{\Ham}   {{\mathcal{H}}}

\newcommand{\kbf}      {\textbf{k}}

\newcommand{\kprbf}      {\textbf{k}^\prime}
\newcommand{\kpr}      {k^\prime}
\newcommand{\qbf}      {\textbf{q}}
\newcommand{\Qbf}      {\textbf{Q}}

\newcommand{\ibf}      {\textbf{i}}
\newcommand{\jbf}      {\textbf{j}}
\newcommand{\rbf}      {\textbf{r}}

\newcommand{\Ugs}{U_{\rm GS}}

\newcommand{\Ub}{U_{\rm GS}}

\newcommand{\Upl}{U_{\rm plrn}}
\newcommand{\Rq}{R_\qbf}
\newcommand{\Rqd}{R_\qbf^\dagger}

\newcommand{\dtau}{\partial_\tau}
\newcommand{\invrN}{\frac1{\sqrt{N}}}
\newcommand{\gred}{g_\qbf}

\begin{document}

\title{Zero-Temperature Phases of the 2D Hubbard-Holstein Model:\\ A Non-Gaussian Exact Diagonalization Study}
\author{Yao Wang}
\email{yaowang@g.harvard.edu}
 \affiliation{Department of Physics, Harvard University, Cambridge, Massachusetts 02138, USA}
 \affiliation{Department of Physics and Astronomy, Clemson University, Clemson, South Carolina 29631, USA}
 \author{Ilya Esterlis}
 \affiliation{Department of Physics, Harvard University, Cambridge, Massachusetts 02138, USA}
 \author{Tao Shi}
 \email{tshi@mail.itp.ac.cn}
 \affiliation{CAS Key Laboratory of Theoretical Physics, Institute of Theoretical Physics, Chinese Academy of Sciences, Beijing 100190, China}
 \affiliation{CAS Center for Excellence in Topological Quantum Computation, University of Chinese Academy of Sciences, Beijing 100049, China}
  \author{J. Ignacio Cirac}
\affiliation{Max-Planck-Institut f\"ur Quantenoptik, Hans-Kopfermann-Strasse. 1, 85748 Garching, Germany}
\affiliation{Munich Center for Quantum Science and Technology (MCQST), Schellingstr. 4, 80799 M\"unchen, Germany}
 \author{Eugene Demler}
 \affiliation{Department of Physics, Harvard University, Cambridge, Massachusetts 02138, USA}
\date{\today}
\begin{abstract}
    We propose a numerical method which embeds the variational non-Gaussian wavefunction approach within exact diagonalization, allowing for efficient treatment of correlated systems with both electron-electron and electron-phonon interactions. Using a generalized polaron transformation, we construct a variational wavefunction that absorbs entanglement between electrons and phonons into a variational non-Gaussian transformation; exact diagonalization is then used to treat the electronic part of the wavefunction exactly, thus taking into account high-order correlation effects beyond the Gaussian level. Keeping the full electronic Hilbert space, the complexity is increased only by a polynomial scaling factor relative to the exact diagonalization calculation for pure electrons. As an example, we use this method to study ground-state properties of the two-dimensional Hubbard-Holstein model, providing evidence for the existence of intervening phases between the spin and charge-ordered states. In particular, we find one of the intervening phases has strong charge susceptibility and binding energy, but is distinct from a charge-density-wave ordered state, while the other intervening phase displays superconductivity at weak couplings. This method, as a general framework, can be extended to treat excited states and dynamics, as well as a wide range of systems with both electron-electron and electron-boson interactions.
\end{abstract}
\maketitle

\section{Introduction}
Strongly correlated systems pose important theoretical questions about the nature of interacting systems at intermediate and strong coupling. Away from weak coupling, traditional mean-field or perturbative approaches often fail to accurately describe the physics, especially in cases with competing and/or intertwined ordering tendencies. 

In the condensed-matter setting, models are commonly classified into interacting electrons, interacting bosons, and interacting electron-boson systems. Advances in unbiased numerical many-body methods, including exact diagonalization (ED)\,\cite{dagotto1994correlated, bonvca1999holstein}, quantum Monte Carlo (QMC)\,\cite{gull2011continuous,rubtsov2005continuous} and density-matrix renormalization group (DMRG)\,\cite{white1992density,schollwock2011density}, have greatly expanded our understanding of the fermionic and bosonic Hubbard models, together with their variants. For example, recent numerical solutions of the single- and three-band Hubbard models have shed light on the stripe and $d$-wave superconducting phases in doped cuprates\,\cite{zheng2016ground,huang2017numerical, zheng2017stripe,ido2018competition,jiang2018superconductivity}. Although the Hubbard model is often considered to be a prototype microscopic model, experimental evidence suggests coupling to phonons can also play an important role in the low-energy physics of correlated materials. For example, STM measurements have shown a significant isotope effect on the second-derivative tunneling current\,\cite{lee2006interplay}; spectral experiments have shown significant lattice effects in cuprates, starting from the underdoped regime\,\cite{shen2004missing}, to optimal\,\cite{lanzara2001evidence} and overdoped regimes\,\cite{he2018rapid}; phonon softening has also been observed using Raman\,\cite{thomsen1988untwinned} and neutron scattering\,\cite{reznik2006electron}. These observations suggest that electron-electron (e-e) and electron-phonon (e-ph) interactions should be taken into account simultaneously in order to properly understand the rich phenomena observed in many correlated materials. 

A significant barrier to understanding the low-energy physics of models with both e-e and e-ph interactions is the challenge they pose to conventional numerical methods. On the one hand, numerical many-body approaches, such as ED and DMRG, have achieved great success in analyzing correlated electronic systems in the past decades. With the improvement of both algorithms and high-performance supercomputers, these approaches not only evaluated the ground state properties precisely, but also calculated the spectroscopies and dynamics in a well-controlled way\,\cite{white2004real,manmana2007strongly,balzer2011krylov}. However, extending efficient numerical techniques to include phonons remains challenging. The bosonic Hilbert space is infinite-dimensional, and the total allowed phonon number has to be truncated to a small value (on the order of 1--5 phonons per site). This has largely limited the study of strongly coupled e-ph systems, \emph{e.g.}~the Peierls charge-density-wave (CDW) systems. 

On the other hand, approximate methods based on variational wavefunctions provide an alternative route to analyze correlated systems. For example, variational Lang-Firsov transformations have been applied to disentangle e-ph systems in the long-wavelength limit\,\cite{fehske1994polaron,chatterjee2004hubbard}. A more intricate Jastrow variational wavefunction has been employed to examine the competing spin, charge, and superconducting orders via a particular mean-field decoupling of the electrons.\,\cite{karakuzu2017superconductivity, ohgoe2017competition} More recently, these variational approaches were generalized to the non-Gaussian class of wavefunctions. With the non-Gaussian transformation chosen to be a generalized polaron transformation, this method gives a good estimate of the e-ph ground state\,\cite{shi2018variational}; with specific parity transformation, this class of wavefunctions also perfectly decouples the Kondo and Anderson models\,\cite{ashida2018solving,ashida2018variational,shi2019ultrafast}. However, extending the method to systems with e-e interactions has not been straightforward, due to the fact that four-fermion interaction terms make the parameter space much more complicated. Besides, effective e-e interactions can be generated when disentangling the e-ph coupling. The absence of quantum fluctuations in the Gaussian state limits the accuracy for even pure e-ph systems. This issue becomes even more crucial for the calculation of dynamics, due to greater complexity of the polaronic dressing\,\cite{wang2019nonGaussian}. More precise treatment of electronic correlations is therefore imperative. 

To combine the merits of these two philosophies, we propose the hybrid non-Gaussian exact diagonalization (NGSED) method. By adding the polaronic non-Gaussian ansatz for the phonon dressing (to be described in more detail below) to the ED-based electronic calculation, we increase the computational complexity only by a polynomial factor. At the same time, the inclusion of the full electronic Hilbert space and many-body wavefunction addresses the fluctuation issue of pure variational approaches, reducing the bias incurred by a mean-field treatment of the correlated electronic state. Similar embedding ideas have been attempted in a few numerical studies. For example, the iterative optimized phonon implementation has been applied to ED\,\cite{weisse2000optimized} and cluster perturbation theory\,\cite{ning2006phonon}. However, even on an optimized basis, the phonon number still spans a huge Hilbert space, limiting the calculations to a six-site chain. The classical phonon approximation\,\cite{dobry1994effects} and the standard Lang-Firsov transformation\,\cite{takada2003possibility} were also employed to disentangle the local interactions in QMC and ED, but ignorance of explicit phonon wavefunctions prevents an accurate description of fluctuations of both effective tunneling and interactions. A very similar idea of embedding Lang-Firsov transformations with ED has been attempted in the $t\!-\!J$ model\,\cite{fehske1995hole,bauml1998optical}. With only a local dressing parameter, these embedding calculations still failed to capture the fluctuations caused by the polaronic dressing. Therefore, a natural extension is the embedding of a variational phonon wavefunction and polaron transformation into an exact numerical technique -- this forms the intuition of our NGSED method.

Although the idea of embedding non-Gaussian transformations with numerical many-body techniques can be extended to a variety of problems, we focus on the e-ph system as a concrete topic in this paper. We introduce the NGSED method for a generic e-ph model and present the iterative approach to evaluate the ground-state properties. To assess the accuracy of the variational wavefunction, we benchmark the method against an exact QMC solution of the Holstein model. We then focus on the Hubbard-Holstein model, where we examine the ground-state properties and their dependence on the e-e and e-ph interactions, phonon energy, and doping. We observe a shift in the antiferromagnetic (AFM) phase boundary, which is explained through the form of effective e-e interactions. We identify a region between AFM and charge-density-wave (CDW) states in which both charge and spin orders are absent. This region can further be divided: one subregion has enhanced charge susceptibility and considerable binding energy, possibly corresponding to a two-dimensional (2D) analog of the Luther-Emery liquid observed in the one-dimensional (1D) Hubbard-Holstein model\,\cite{Greitemann2015arxiv}; the other subregion exhibits superconductivity at the weak-coupling side but gradually becomes metallic for stronger coupling. In contrast to the conclusions obtained using pure variational wavefuctions\,\cite{karakuzu2017superconductivity, ohgoe2017competition}, we do not see a dramatic broadening of the superconducting phase on the weak-coupling side, consistent with unbiased QMC results\,\cite{hohenadler2019dominant}. Complementing previous high-temperature QMC studies, truncated-phonon ED studies, and zero-temperature variational studies, this work sheds new light on the phases in such a competing-order system. 

The organization of this paper is as follows. We first introduce the NGSED method and relevant derivations in Sec.~\ref{sec:method}. Then we apply it to the Holstein and Hubbard-Holstein models and discuss the ground-state properties in Sec.~\ref{sec:HHModel}. We conclude our method and simulations in Sec.~\ref{sec:conclusion}, together with the outlook of this method for other systems.

\section{Model and Derivations}\label{sec:method}
We present the derivation of relevant formulas for a generic electron-phonon system in this section, before focusing on the Hubbard-Holstein model with specific form of e-e and e-ph interactions. A generic electron-phonon model can be expressed by the Hamiltonian
\begin{eqnarray}\label{eq:genericephHam}
\Ham = \sum_{\kbf\sigma} (\varepsilon_\kbf -\mu)c^\dagger_{\kbf\sigma}c_{\kbf\sigma} + \Ham_{e-e} + \Ham_{e-\mathrm{ph}}+ \Ham_{\rm ph} 
\end{eqnarray}
where $c_{\kbf\sigma}$ ($c_{\kbf\sigma}^\dagger$) annihilates (creates) an electron at momentum $\kbf$ with spin $\sigma$, with a dispersion relation $\varepsilon_\kbf$ and chemical potential $\mu$. $N$ is the overall site number. Within second quantization, $c_{\kbf\sigma}$ takes the reciprocal representation with respect to the annihilation operator of the Wannier orbital $c_{\kbf\sigma} =  \sum_{\ibf} e^{-i\kbf\cdot \rbf_\ibf} c_{\ibf\sigma}/\sqrt{N}$. Apart from the bare dispersion, the $\Ham_{e-e}$, $\Ham_{e-\mathrm{ph}}$ and $\Ham_{\rm ph} $ terms represent the contributions from e-e interactions, e-ph coupling and phonon energy, respectively. 

In general, the phonon part of Hamiltonian is
\begin{eqnarray}
\Ham_{\rm ph} = \sum_\qbf \omega_\qbf a_\qbf^\dagger a_\qbf =\frac14\sum_\qbf \omega_\qbf\Rqd \Rq\ ,
\end{eqnarray}
and the e-ph coupling part is\,\cite{genericEPC}
\begin{eqnarray}
\Ham_{e-\mathrm{ph}}=  \frac1{\sqrt{N}} \sum_\qbf g_\qbf (a_{\qbf}+a_{-\qbf}^{\dagger }) \rho_\qbf\,.
\end{eqnarray}
Here, the $\omega_\qbf$ describes the phonon dispersion, $g_\qbf$ parametrizes the e-ph interaction at a wavevector $\qbf$; $a_\qbf$ annihilates a phonon at momentum $\qbf$ and $\rho_{\qbf}=\sum_{\ibf\sigma} n_{\ibf\sigma}e^{-i\qbf\cdot \rbf_\ibf}$ is the electron density.
For convenience in subsequent derivations,
we employ the bosonic quadrature notation $\Rq =( x_\qbf, p_\qbf )^T $, with the canonical position $x_{\qbf}=a_{\qbf}+a_{-\qbf}^{\dagger }$ and momentum $p_{\qbf}=i( a_{-\qbf}^{\dagger}-a_{\qbf}) $ determined by the phonon annihilation operator. These canonical operators fulfill the commutation relations
\begin{eqnarray}
    \Big[x_\qbf^\dagger,p_{\qbf^\prime}\Big]=\left[ x_{\qbf},p_{\qbf^\prime}^\dagger\right] =2i\delta _{\qbf,\qbf^\prime}\,.
\end{eqnarray}
Thus, the parts of Hamiltonian relevant to phonons can be rewritten as
\begin{eqnarray}
\Ham_{e-\mathrm{ph}}+\Ham_{\rm ph} =\frac1{\sqrt{N}}\sum_\qbf \gred \Rqd e_1 \rho_\qbf+\frac14\sum_\qbf \omega_\qbf\Rqd \Rq,
\end{eqnarray}
with $e_1 = (1,0)^T$.

Without loss of generality, we allow the parameters $g_\qbf$ and $\omega_\qbf$ to vary over momentum space, but obeying the time-reversal symmetry, i.e., $g_\qbf=g_{-\qbf}$ and $\omega_\qbf = \omega_{-\qbf}$. For the electron interaction part $\Ham_{e-e}$, the only restriction we place is that it commutes with the local density operators $n_{\ibf}$. Thus we allow for any combination of density and spin operators, such as the on-site Hubbard or long-range Coulomb interactions.

To describe the e-ph entangled system in the simplest form, we consider the wavefunction ansatz
\begin{eqnarray}\label{eq:wvfuncansatzFull}
    \big|\Psi\big\rangle  = \Upl |\psi_{\rm ph}\rangle \otimes|\psi_{\rm e}\rangle.
\end{eqnarray}
Here, the right-hand-side is a direct product of electron and phonon states, where $|\psi_{\rm e}\rangle$ is treated as a full many-body state while $ |\psi_{\rm ph}\rangle$ is a coherent Gaussian state
\begin{eqnarray}\label{eq:wvfuncansatzPh}
    |\psi_{\rm ph}\rangle = e^{-\frac12 R_0^T \sigma_y \Delta_R} e^{-i\frac14 \sum_\qbf \Rqd \xi_\qbf  \Rq} |0\rangle = \Ub |0\rangle,
\end{eqnarray}
in which $\sigma_y$ is the Pauli matrix.
The polaron transformation $\Upl$ creates entanglement between these two parts of the wavefunction:
\begin{eqnarray}\label{eq:nonGSansatz}
    \Upl = e^{i\frac1{\sqrt{N}}\sum_\qbf \lambda_\qbf p_{-\qbf}. \rho_\qbf}
\end{eqnarray}
In the above wavefunction prototype, the $\Delta_R$, $\xi_\qbf$ and $\lambda_\qbf$ are variational parameters. An important feature of the wavefunction ansatz in Eq.~\eqref{eq:wvfuncansatzFull} is that this wavefunction gives exact solutions to the e-ph problem in both the adiabatic ($\omega=0$) and anti-andiabatic ($\omega=\infty$) limits. In the adiabatic limit, phonons can be treated as a classical field, mean-field theory becomes exact, and the Gaussian wavefunction gives an exact description of the phonon. Thus, the system becomes pure electronic and can be precisely solved by ED. In the anti-adiabatic limit, the phonon field can be integrated out, yielding an instantaneous, attractive on-site interaction; \emph{i.e.} the attractive Hubbard model. The ED step again solves this problem exactly. As it is exact in both the adiabatic and anti-adiabatic limits, we expect Eq.~\eqref{eq:wvfuncansatzFull} does not induce significant bias in realistic models with finite $\omega$. The accuracy of this assumption will be further assessed through the comparison with exact DQMC solutions [see Sec.~\ref{sec:Holstein}]. In contrast, a Gaussian ansatz for the fermionic wavefunction would not accurately describe the system in either limits, because the quantum fluctuations become important with the presence of electronic interactions in $\Ham_{e-e}$.

Note, in principle, the coherent part of the phonon wavefunction $|\psi_{\rm ph}\rangle$ can involve displacements for all different momenta. However, any finite value of finite-$\qbf$ displacement would lead to the explicit breaking of translational symmetry and over-estimate the tendency of charge ordering. Therefore, to avoid possible biases induced by the symmetry-breaking Gaussian states, we neglect any finite-$\qbf$ displacements in Eq.~\eqref{eq:wvfuncansatzPh}.
Physically, it means phonons cannot really condense at a finite momentum, though the system might exhibit dramatically enhanced fluctuations. We impose this strong assumption because spontaneous symmetry breaking is not possible in such a small cluster. Therefore, to fairly study the competition between spin- and charge-density-wave states, we only discuss their susceptibilities rather than long-range ordered states. This assumption also highly reduces the Hilbert space dimension due to the momentum conservation.

The above polaron transformation generalizes the Lang-Firsov transformation\,\cite{lang1962}. Historically, the Lang-Firsov transformation has been widely exploited in electron-boson systems to disentangle the coupling and simplify the calculation. To tackle the Hubbard-Holstein model, early attempts have extended it to a variational transformation\,\cite{fehske1994polaron,chatterjee2004hubbard}. These transformations have shown advantages in solving the Holstein model\,\cite{das2008thermodynamic}, Hubbard-Holstein model\,\cite{sankar2016quantum,ghosh2018study}, Anderson-Holstein model\,\cite{raju2015effect,monisha2016persistent}, and anharmonic phonons\,\cite{lavanya2017metallicity}. However, due to the limitation of the numerical treatment on either the phonon or electronic side, these variational transformations were restricted only to a $\qbf-$independent $\lambda_\qbf$. This treatment ignores the longer-range spatial fluctuation of the effective interaction mediated by the phonon, which we will show plays a significant role near the quantum phase transition. A direct consequence of this simplification is the overestimation of the CDW instability [we will further discuss this in Sec.~\ref{sec:HHModel}]. This limitation necessitates the generalization of this transformation to a polaronic non-Gaussian transformation in Eq.~\eqref{eq:nonGSansatz}, where $\lambda_\qbf$ is allowed to vary for different momenta.

By constructing the wavefunction through Eq.~\eqref{eq:wvfuncansatzFull}, we can evaluate the ground state with the manifold spanned by the variational parameters and the many-body electronic wavefunctions. Variational parameters are determined by minimizing the energy
\begin{eqnarray}\label{eq:energytarget}
    E\Big(\{\lambda_\qbf\},\Delta_R,\{\xi_\qbf\},|\psi_e\rangle\Big) =  \big\langle\Psi\big|\Ham \big|\Psi\big\rangle\,.
\end{eqnarray}
Numerically, the optimization can be iteratively achieved by decomposing into the electronic and bosonic state, with coupled coefficients. Each of them can be treated as a correction to the effective Hamiltonian while optimizing the other. Thus, for an equilibrium state, we minimize the total energy along two gradient directions sequentially. With an initial guess not far from the global minimum, we expect the many-body electronic state and variational phonon state to converge to the ground state self-consistently. In the following two subsections, we describe the procedures for evolving these two parts of the state. Afterward, we describe the above self-consistent iteration in a more strict manner using notations introduced in these two subsections.

\subsection{Electron Ground State: Exact Diagonalization}\label{sec:electronOptimization}
We first optimize the electronic state (minimizing the energy), keeping fixed the variational parameters in $\Upl$ and $\Ugs$. Then, Eq.~\eqref{eq:energytarget} becomes an unrestricted minimization of energy
\begin{eqnarray}
    E(|\psi_e\rangle) =  \big\langle\psi_e\big|\Ham_{\rm eff}\big|\psi_e\big\rangle
\end{eqnarray}
in the full electronic Hilbert space, where the effective electronic Hamiltonian is given by tracing over the phonon state
\begin{eqnarray}
    \Ham_{\rm eff} = \langle\psi_{\rm ph} | \Upl^\dagger\Ham \Upl|\psi_{\rm ph}\rangle .
\end{eqnarray}
The $\Ham_{\rm eff}$ is an operator only on the electronic Hilbert space. Since the phonon state is Gaussian, the expression for $\Ham_{\rm eff}$ can be obtained analytically:
\begin{eqnarray}\label{eq:effElectronHam}
    &&-t\!\sum_{\jbf\sigma\alpha\delta_\alpha}\! \left\langle\! e^{i\frac1{\sqrt{N}}\!\sum_\qbf\! \lambda_\qbf\Rqd S_\qbf^\dagger e_2 e^{-i\qbf\cdot\jbf}\left(1-e^{-i\qbf\cdot\mathbf{\delta}_\alpha}\right) }\!\right\rangle_0 \!c_{\jbf+\mathbf{\delta}_\alpha,\sigma}^\dagger c_{\jbf\sigma}\nonumber\\
    && + \!\frac14\!\sum_\qbf\!\omega_\qbf\! \left\langle\left(\!\Rqd S_\qbf^\dagger\!+\!\Delta_R^T\delta_{\qbf 0}\!\right)\!\left(\!S_\qbf\Rq\!+\!\Delta_R\delta_{\qbf 0}\!\right)\right\rangle_0\! - \sum_\qbf\frac{\omega_\qbf}{2}\nonumber\\
&&-\invrN\sum_\qbf\left(\lambda_\qbf\omega_\qbf - \gred\right)\left(\left\langle\Rqd\right\rangle_0 S_\qbf^\dagger +\Delta_R^T\delta_{\qbf 0}\right)  e_1\rho_\qbf\nonumber\\
&& -\frac1{2N}\sum_\qbf\sum_{\kbf,\kbf^\prime\atop \sigma, \sigma^\prime} V_\qbf\, c_{\kbf+\qbf,\sigma}^\dagger c_{\kbf\sigma} c_{\kbf^\prime-\qbf, \sigma^\prime}^\dagger c_{\kbf^\prime \sigma^\prime}+ \Ham_{e-e}\nonumber\\
&=& \sum_{\kbf\sigma}\tilde{\varepsilon}_\kbf  n_{\kbf\sigma}+ \frac14\sum_\qbf\omega_\qbf\left(\Tr [ \Gamma_\qbf]-2\right)\nonumber\\
&& -\frac1{2N}\sum_\qbf\sum_{\kbf,\kbf^\prime\atop \sigma, \sigma^\prime} V_\qbf\, c_{\kbf+\qbf,\sigma}^\dagger c_{\kbf\sigma} c_{\kbf^\prime-\qbf, \sigma^\prime}^\dagger c_{\kbf^\prime \sigma^\prime} + \Ham_{e-e}\nonumber\\
&& +\frac14\Delta_R^T \omega_0 \Delta_R +\frac1{\sqrt{N}} (g_0-\lambda_0 \omega_0)\Delta_R^T e_1 \rho_0
\end{eqnarray}
Here the variational parameters for the phonon state are rewritten as $\Gamma_\qbf = S_\qbf S_\qbf^\dagger$, with the matrix $S_\qbf$ representing the linearization of the $\Ub$, \emph{i.e.}, $\Ub^\dagger \Rq\Ub = S_\qbf \Rq$.

The polaronic dressing is reflected in the effective kinetic energy, \emph{i.e.}, the renormalized band dispersion $\tilde{\varepsilon}_\kbf = -2t_x \cos k_x - 2t_y \cos k_y - \mu$, where
\begin{eqnarray}\label{eq:effHopping}
    t_\alpha = te^{-\sum_\qbf\frac{|\lambda_\qbf|^2}{N}(1-\cos q_\alpha) e_2^T \Gamma_\qbf e_2},
\end{eqnarray}
and the effective electronic attraction
\begin{eqnarray}\label{eq:effInteraction}
    V_\qbf = 4g_\qbf \real[\lambda_\qbf] - 2 \omega_\qbf|\lambda_\qbf|^2.
\end{eqnarray}
In the above derivations, we have employed the assumption that $[\Ham_{e-e},n_{\ibf}]=0$. Note that in the last step of Eq.~\eqref{eq:effElectronHam}, the electron density at momentum $\qbf=0$ is nothing but the total occupation $N_e$ in a micro-canonical ensemble. Therefore, the energy minimization with respect to $\Delta_R$ can be done immediately, leading to $\Delta_R = (2N_e(\lambda_0\omega_0 - g_0)/\sqrt{N}\omega_0, 0)^T$. As will be shown later in Eq.~\eqref{eq:imagEOMlambda}, $\lambda_0= g_0/\omega_0 $ for the saddle-point solution. Therefore, for the purpose of calculating the ground state, it is convenient to set $\Delta_R\equiv 0$.

Different from the original Lang-Firsov transformation in the atomic limit, both the kinetic and interaction energies in the effective electronic Hamiltonian are renormalized by the phonons. The variational parameters allow us to find a balance between these two effects and minimize the entanglement between electrons and phonons by optimizing $\lambda_\qbf$\,\cite{takada2003possibility}. Moreover, different from the widely used modified Lang-Firsov transformations\,\cite{fehske1994polaron,chatterjee2004hubbard}, the generalized polaron transformation and phonon Gaussian state naturally give momentum fluctuations of the effective interaction $V_\qbf$. In later discussions, we will show that this fluctuation is crucial near the phase boundary.

Since we keep the full electronic Hilbert space, it is straightforward to diagonalize the matrix $\Ham_{\rm eff}$ and find the ground state through a standard Lanczos approach. As we will discuss below, the ground state can be obtained alternatively through a flow equation -- imaginary time evolution. However, with the full Hilbert space information, computing a matrix diagonalization is much cheaper than performing a time evolution, though the latter has been widely used in variational approaches.

\subsection{Phonon Ground State: Imaginary Time Evolution}
Keeping the electronic wavefunction fixed, the energy minimization in Eq.~\eqref{eq:energytarget} can be achieved through the imaginary time evolution
\begin{eqnarray}
    \dtau|\Psi(\tau)\rangle\! =\! -\Big(\Ham - \big\langle\Psi(\tau)\big|\Ham \big|\Psi(\tau)\big\rangle\Big) |\Psi(\tau)\rangle\,.
\end{eqnarray}
Restricting this equation to the variational class of states, one has to project the right-hand side (RHS) on the tangential plane (see the derivations below). This procedure guarantees the monotonic decrease of energy while maintaining the normalization of the wavefunction. If we restrict $\Delta_R=0$ as mentioned above, the derivative of the variational wavefunction becomes
\begin{widetext}
\begin{eqnarray}\label{eq:imagdtauphi}
\dtau\big |\Psi(\tau)\big\rangle =\Upl \Ugs\Big[- \frac14\sum_\qbf\Rqd S_\qbf^\dagger \sigma_y\dtau S_\qbf\Rq  +i\frac1{\sqrt{N}}\sum_\qbf \Rqd S_\qbf^\dagger e_2 \rho_\qbf\dtau\lambda_\qbf \Big]|0_{\rm ph}\rangle \otimes|\psi_{\rm e}\rangle.
\end{eqnarray}
Taking into account the orthogonality of the electronic wavefunction basis, the tangential vectors are $a_\qbf^\dagger a_{-\qbf}^\dagger|0_{\rm ph}\rangle\otimes|\psi_{\rm e}\rangle$ and $a_\qbf^\dagger |0_{\rm ph}\rangle\otimes \rho_\qbf|\psi_{\rm e}\rangle$.\,\cite{zerothOrderTVec} The rotated Hamiltonian is
\begin{eqnarray}\label{eq:imagH}
\Ugs^\dagger\Upl^\dagger\!\Ham\Upl\Ugs\!
&=&  \frac14\omega_\qbf \Rqd S_\qbf^\dagger S_\qbf\Rq-\invrN\sum_\qbf\left(\lambda_\qbf\omega_\qbf - \gred\right)\Rqd S_\qbf^\dagger  e_1\rho_\qbf\nonumber\\
&&-t\!\sum_{\jbf\sigma\alpha\delta_\alpha}\! e^{i\frac1{\sqrt{N}}\sum_\qbf\! \lambda_\qbf\Rqd S_\qbf^\dagger e_2 e^{-i\qbf\cdot\jbf}\left(1-e^{-i\qbf\cdot\delta_\alpha}\!\right) } c_{\jbf+\delta_\alpha,\sigma}^\dagger \!c_{\jbf\sigma} -\frac1{2N}\!\sum_\qbf\sum_{\kbf,\kbf^\prime\atop \sigma, \sigma^\prime}\! V_\qbf\, c_{\kbf+\qbf,\sigma}^\dagger c_{\kbf\sigma} c_{\kbf^\prime-\qbf, \sigma^\prime}^\dagger c_{\kbf^\prime \!\sigma^\prime} \,,
\end{eqnarray}
\end{widetext}
where $\alpha = x, y$ denotes the direction, while $\delta_\alpha$ is a unit vector along the $\alpha$-direction. To determine the evolution of the variational wavefunction, we project the rotated Hamiltonian to the above two sets of tangential vectors.

On the one hand, the projection with respect to the second-order bosonic terms is\,\cite{shi2018variational, wang2019nonGaussian}
\begin{eqnarray}\label{eq:imagsecondBosonDynm}
(1,i) S^\dagger_q\sigma_y\dtau S_\qbf \left(\begin{matrix} 1\\ i\end{matrix}\right) 
&=& (1,i)S_\qbf^\dagger \tilde{\Omega}_\qbf S_\qbf\left(\begin{matrix} 1\\ i\end{matrix}\right) \,.
\end{eqnarray}
The renormalized phonon energy matrix is
\begin{eqnarray}\label{eq:renormalizedPhononFreq}
\tilde{\Omega}_\qbf= \omega_\qbf+\frac{8|\lambda_\qbf|^2}{N}\sum_{k\alpha}t_\alpha \big[1-\cos q_\alpha\big]\langle n_k\rangle\cos k_\alpha E_{22}.
\end{eqnarray}
Here, $E_{22} = e_2 e_2^T $ and $e_2 = (0,1)^T$. Transforming the scalar equation of motion Eq.~\eqref{eq:imagsecondBosonDynm} into a matrix form, we should fill up the missing matrix elements in an anit-Hermitian way, which gives
\begin{eqnarray}
\dtau S_\qbf = \frac12\left[\sigma_y \tilde{\Omega}_\qbf S_\qbf\sigma_y - \Gamma_\qbf \tilde{\Omega}_\qbf S_\qbf \right]\,.
\end{eqnarray}
Absorbing the gauge freedom, we have
\begin{eqnarray}\label{eq:imagEOMBoson}
\dtau \Gamma_\qbf = \sigma_y \tilde{\Omega}_\qbf \sigma_y - \Gamma_\qbf \tilde{\Omega}_\qbf \Gamma_\qbf \,.
\end{eqnarray}

On the other hand, the projection to the other tangential vector gives
\begin{eqnarray}\label{eq:cubicProjection}
&&\dtau \lambda_\qbf (i,-1)S_\qbf^\dagger e_2 \Big\langle\!{\rho_{-\qbf}}{\rho_{\qbf}} \!\Big\rangle\! \nonumber\\
&=&\! 2i\lambda_\qbf (1,i)S_\qbf^\dagger e_2\!\sum_{k\sigma\alpha} \!t_\alpha [\cos k_\alpha\! -\! \cos(k_\alpha\!+\!q_\alpha)]    \Big\langle \!{\rho_{-\qbf}}{c}_{k\sigma}^\dagger {c}_{k\!+\!q,\sigma} \!\Big\rangle\! \nonumber\\
&&+\! \left(\omega_\qbf\lambda_\qbf\! -\!\gred\right)  (1,i) S_\qbf^\dagger e_1 \Big\langle\!{\rho_{-\qbf}}{\rho_{\qbf}} \!\Big\rangle.
\end{eqnarray}
We define the modulated electronic correlation function
\begin{eqnarray}\label{eq:defCorrelations}
\sum_{\kbf\sigma\alpha } \!t_\alpha [\cos k_\alpha \!-\! \cos(k_\alpha\!+\!q_\alpha)]    \Big\langle \!{\rho_{-\qbf}}{c}_{\kbf\sigma}^\dagger {c}_{\kbf\!+\!\qbf,\sigma} \!\Big\rangle \! =  \!\Pi_\qbf \!+  \!i\Theta_\qbf,
\end{eqnarray}
where both $\Pi_\qbf$ and $\Theta_\qbf$ are real-valued functions. Comparing the real and imaginary parts of Eq.~\eqref{eq:cubicProjection}, we have the equation of motion
\begin{eqnarray}\label{eq:imagEOMlambda}
\dtau \lambda_\qbf 
 \!= \!\left(\gred  \!- \! \omega_\qbf\lambda_\qbf\right)  \!e_1^T \Gamma_\qbf e_1 \! 
+ \! 2 \lambda_\qbf  \frac{\Pi_\qbf}{C_\qbf} \!+ \!2 \lambda_\qbf  \frac{\Theta_\qbf}{C_\qbf}e_1^T \Gamma_\qbf e_2
\end{eqnarray}
where the density correlation is $C_\qbf =\langle{\rho_{-\qbf}}{\rho_{\qbf}} \rangle$.

By solving the imaginary-time equations of motion \eqref{eq:imagEOMBoson} and \eqref{eq:imagEOMlambda}, one obtains the variational parameters that minimize the energy for given electronic state $|\psi_e\rangle$. In particular for $\qbf=0$, the renormalized phonon energy in Eq.~\eqref{eq:renormalizedPhononFreq} reduces to $\omega_0$ and the electronic correlations in Eq.~\eqref{eq:defCorrelations} vanish, leading to the saddle-point solution $\lambda_0 = g_0/\omega_0$. This condition has been exploited above to simplify the effective electronic Hamiltonian.

\subsection{Non-Gaussian Exact Diagonalization Iterations}\label{sec:method_NGSiterations}
\begin{figure}[!t]
\begin{center}
\includegraphics[width=\columnwidth]{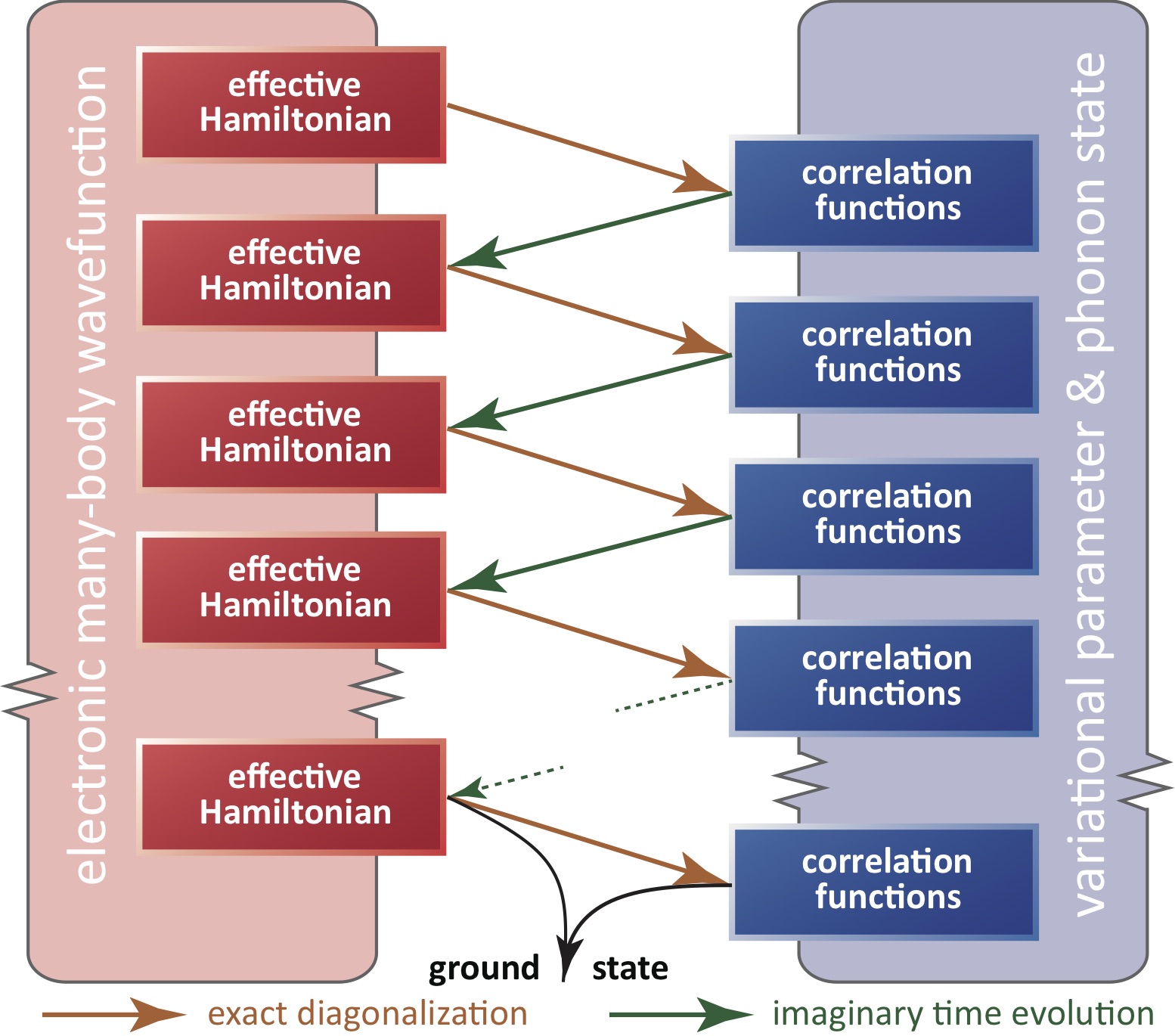}
\caption{\label{fig:cartoon} Schematic illustration of the NGSED iterations towards the ground state of an e-ph system.}
\end{center}
\end{figure}

The above two subsections outline the approach to obtain the electronic ground state with fixed variational parameters, and the ground state of variational wavefunctions with the fixed electronic state. Since the energy minimization is restricted at each step, a global ground state can be obtained only through iterations. Thus, the non-Gaussian exact diagonalization algorithm works as follows:
\begin{enumerate}
    \item Set the initial values of the variational parameters \{$\Gamma_\qbf$\} and \{$\lambda_\qbf$\}. 
    \item Calculate the effective hopping \{$t_\alpha$\} using Eq.~\eqref{eq:effHopping} and effective electronic interactions \{$V_\qbf$\} using Eq.~\eqref{eq:effInteraction}. 
    \item Construct the effective electronic Hamiltonian in Eq.~\eqref{eq:effElectronHam} and perform exact diagonalization to obtain the ($i$-th iteration) electronic ground state $|\psi_e^{(i)}\rangle$.
    \item Based on the electronic many-body wavefunction $|\psi_e^{(i)}\rangle$, calculate the renormalized phonon energy matrix \{$\tilde{\Omega}_\qbf$\} using Eq.~\eqref{eq:renormalizedPhononFreq} and the correlation functions $C_\qbf$, $\Pi_\qbf$ and $\Theta_\qbf$ using Eq.~\eqref{eq:defCorrelations}.
    \item Perform the imaginary time evolution of the variational wavefunction $|\psi_{\rm ph}^{(i)}\rangle$ and the polaronic transformation $\Upl^{(i)}$ using Eqs.~\eqref{eq:imagEOMBoson} and \eqref{eq:imagEOMlambda}.
    \item Repeat 2-5 until the variational parameters \{$\Gamma_\qbf$\} and \{$\lambda_\qbf$\} converge.
\end{enumerate}
The above process is sketched in Fig.~\ref{fig:cartoon}.

\begin{figure}[!t]
\begin{center}
\includegraphics[width=\columnwidth]{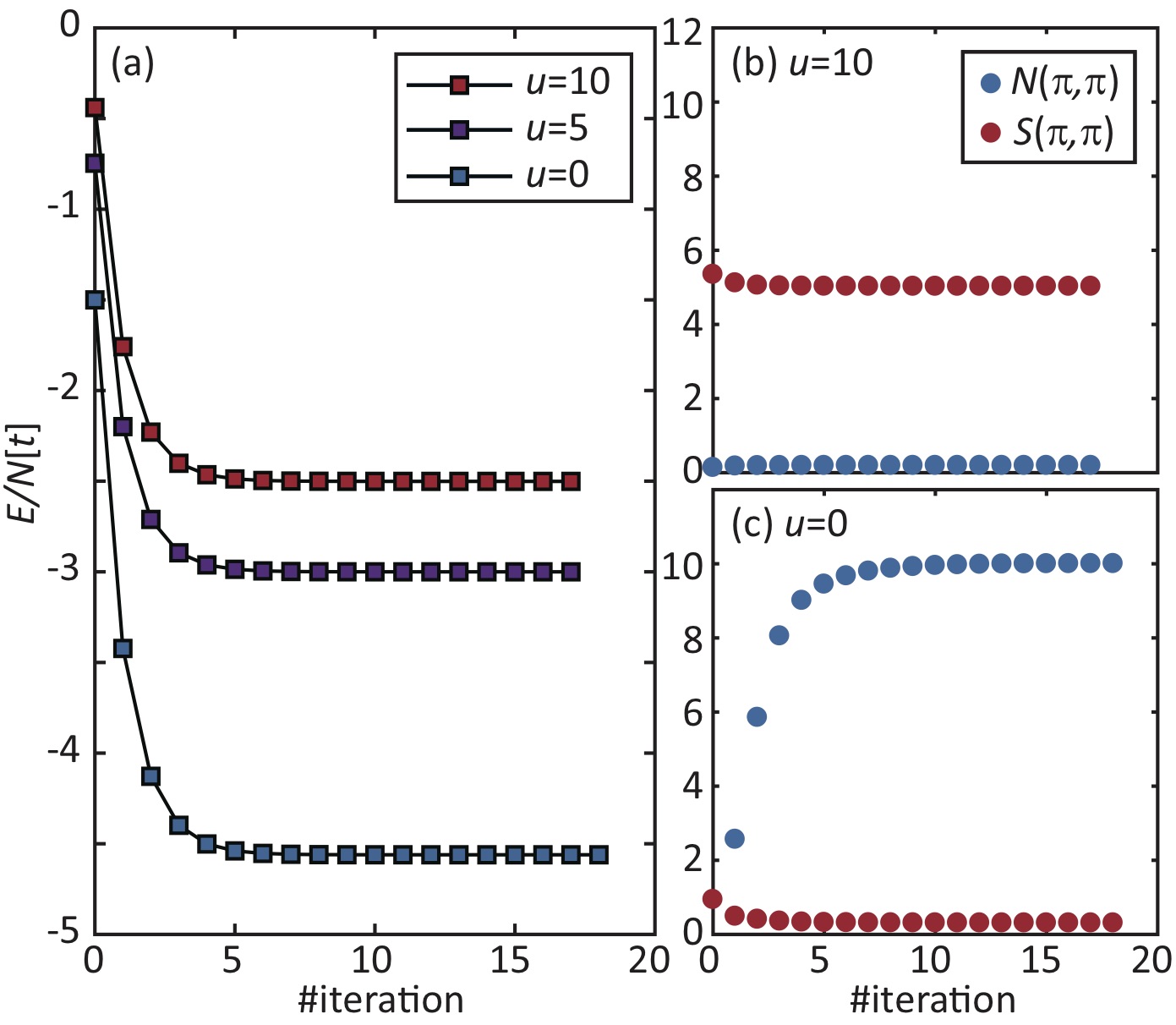}
\caption{\label{fig:evolution}(a) The evolution of site-averaged ground-state energy during the NGSED iterations, for $u=10$, 5 and 0, respectively. (b,c) The evolution of charge (blue) and spin (red) structure factor at the nesting momentum $(\pi,\pi)$ for (b) $u=10$ and (c) $u=0$. All calculations in this figure are obtained for $\lambda=2$ and $\omega=5$.}
\end{center}
\end{figure}

Before we discuss specific parameters, we would like to briefly present an example of the NGSED iterations to give an overview of how the ground state is obtained.  Figure~\ref{fig:evolution}(a) shows the evolution of the energy per site ($E/N$) during the iteration for the Hubbard-Holstein model with $\lambda=2$ and $\omega=5$. The model and model parameters ($u$, $\lambda$ and $\omega$) will be introduced and discussed later in Sec.~\ref{sec:HHModel}. For all three $u$ values, the energy drops rapidly in the first five iterations and starts to saturate. For this sets of model parameters, it takes $\sim30$ iterations to converge with an accuracy of $10^{-6}$. 

To analyze the ordering tendencies of the many-body state, we evaluate the charge structure factor $N(\qbf) = \langle \rho_{-\qbf}\rho_\qbf\rangle /N$ and spin structure factor
\begin{eqnarray}
    S(\qbf) = \frac1{N} \sum_{\kbf,\sigma}\sum_{\kbf^\prime,\sigma^\prime} \sigma^\prime \sigma \left\langle  c_{\kbf^\prime+\qbf\sigma^\prime}^\dagger c_{\kbf^\prime\sigma^\prime}  c_{\kbf-\qbf\sigma}^\dagger c_{\kbf\sigma}\right\rangle.
\end{eqnarray}
These structure factors reflect the charge and spin ordering tendencies at certain momenta. The evolution of these observables as a function of iteration number, at the nesting momentum $\qbf=(\pi,\pi)$, is shown in Figs.~\ref{fig:evolution}(b) and (c). As the variational parameters converge, observables adjust to reflect the ordering tendencies determined by the model parameters. We will discuss the detailed parameter dependence and momentum dependence of observables in Sec.~\ref{sec:HHModel}.

\section{Equilibrium Properties of the Hubbard-Holstein Model}\label{sec:HHModel}
In this section, we apply the NGSED approach to a specific strongly correlated e-ph model and study the equilibrium properties. A typical model describing correlated electrons and phonons is the Hubbard-Holstein (HH) model\,\cite{Hubbard,Holstein}, whose Hamiltonian is
\begin{eqnarray}
    \Ham_{\rm HH}&=& -t\sum_{\langle \ibf,\jbf\rangle,\sigma}  \left[c_{\jbf\sigma}^\dagger c_{\ibf\sigma} + h.c.\right] + U\sum_\ibf n_{\ibf\uparrow}n_{\ibf\downarrow} \nonumber\\
    && + \frac{g}{\sqrt{N}}\sum_{\kbf,\qbf,\sigma} \! x_\qbf c_{\kbf\sigma}^\dagger c_{\kbf+\qbf,\sigma} +\omega\sum_\qbf a_\qbf^\dagger a_\qbf.
\end{eqnarray}
The HH model is a particular example of the generic e-ph system in Eq.~\eqref{eq:genericephHam}. Here, we only consider the nearest-neighbor electron hopping parametrized by the integral $t$, and on-site Hubbard interaction $U$. Both the electron-phonon coupling $g$ and the phonon energy $\omega$ are restricted to be momentum-independent in the HH model. In this case, one can define the dimensionless \textit{e-e} and \textit{e-ph} coupling strengths $u=U/t$ and $\lambda=g^2/t\omega$, respectively. Note, this $\lambda$ is distinct from the variational parameters $\lambda_\qbf$. We adopt this notation as it is standard in the HH model and non-Gaussian literature.

The equilibrium phases of the Hubbard-Holstein model have been studied using different methods. Early studies have examined the equilibrium properties of the 1D HH model, using ED with optimized phonon basis\,\cite{weisse2000optimized, fehske2002peierls}, local Lang-Firsov transformation\,\cite{takada2003possibility}, QMC\,\cite{clay2005intermediate,hohenadler2013excitation,Greitemann2015arxiv,hohenadler2018density}, DMRG\,\cite{tezuka2005density,tezuka2007phase,fehske2008metallicity}, cluster perturbation theory\,\cite{ning2006phonon}, and density-matrix embedding method\,\cite{sandhoefer2016density}. The common results indicated CDW/AFM competition on either side of the anti-adiabatic limit $u=2\lambda$ and an intermediate regime between the ordered phases. This intermediate regime was originally claimed to be superconducting\,\cite{clay2005intermediate}, but more recently confirmed to be a Luther-Emery liquid with quasi-long-range charge and superconducting correlations\,\cite{Greitemann2015arxiv}. The other extreme limit of infinite dimensions has been studied extensively using DMFT\,\cite{bauer2010competing,murakami2013ordered}. These studies indicate the absence of an intermediate phase.

In the context of correlated high-$T_c$ materials, the study of two-dimensional systems is more relevant. However, due to the limitations of numerical techniques, the study of the 2D Hubbard-Holstein model is relatively rare. Using determinant QMC (DQMC), Nowadnick \emph{et al.}~studied the phase diagram of the 2D Hubbard-Holstein model at high temperature (lower temperatures being restricted by the fermion-sign problem) and characterized the metallic phase between the competing ordered phases\,\cite{nowadnick2012competition,nowadnick2015renormalization}. These studies were followed by ED studies at zero temperature. However, due to the infinite Hilbert-space dimensions, these ED studies of HH model were restricted to a one-phonon truncation at small clusters\,\cite{nath2016phonon,nath2015interplay} or single-phonon-mode simplification\,\cite{payeur2011variational, wang2016using, wang2018light}. These calculations, though also exact, are highly restricted by the coupling strength and fillings due to the model simplification. The variational local Lang-Firsov transformation was also applied to 2D ($t$-$J$)-Holstein models, with either Gutzwiller approximation\,\cite{sankar2014self} or exact treatment of the electrons. As mentioned above, this transformation is already close to the generic polaron transformation employed in this work, but the ignorance of the spatial fluctuations makes crucial differences in this context. More recently, the phases of the 2D Hubbard-Holstein model were examined using variational Monte Carlo (VMC), where an $s$-wave superconducting phase was identified in the weak-coupling limit\,\cite{karakuzu2017superconductivity, ohgoe2017competition}. However, the nature of the variational wavefunction biased the system toward superconductivity and the results have been challenged by unbiased QMC studies\,\cite{hohenadler2019dominant}.

Thus, the 2D HH model provides a good platform to demonstrate the capability of the new NGSED method, due to both its known physical properties in the $u=0$ and $\lambda=0$ limits, and important open questions regarding the phase diagram, especially the existence and nature of intermediate phases, which have been challenging to address by the methods in previous studies. With the NGSED approach, we push the ED calculation to a relatively large cluster -- a $4\times4$ system, where vital high-symmetry momenta are included. Although the phonon part of the wavefunction is variational, we minimize bias by treating the electronic part as a full many-body wavefunction. We benchmark the method by comparing with DQMC in a parameter regime where the fermion-sign problem is absent. We use the parallel Arnoldi method\,\cite{lehoucq1998arpack} to determine the ground state wavefunction and the Runge-Kutta Dormand-Prince 5 method to solve the imaginary time evolution. In the following subsections, we first benchmark the NGSED method in the Holstein model with only e-ph interaction. Then we discuss the ground-state properties of the half-filled Hubbard-Holstein model at a fixed phonon frequency. We conclude by briefly examining the impact of phonon frequency and carrier doping in this system.

\subsection{The $u=0$ Limit: the Holstein Model}\label{sec:Holstein}
\begin{figure}[!t]
\begin{center}
\includegraphics[width=\columnwidth]{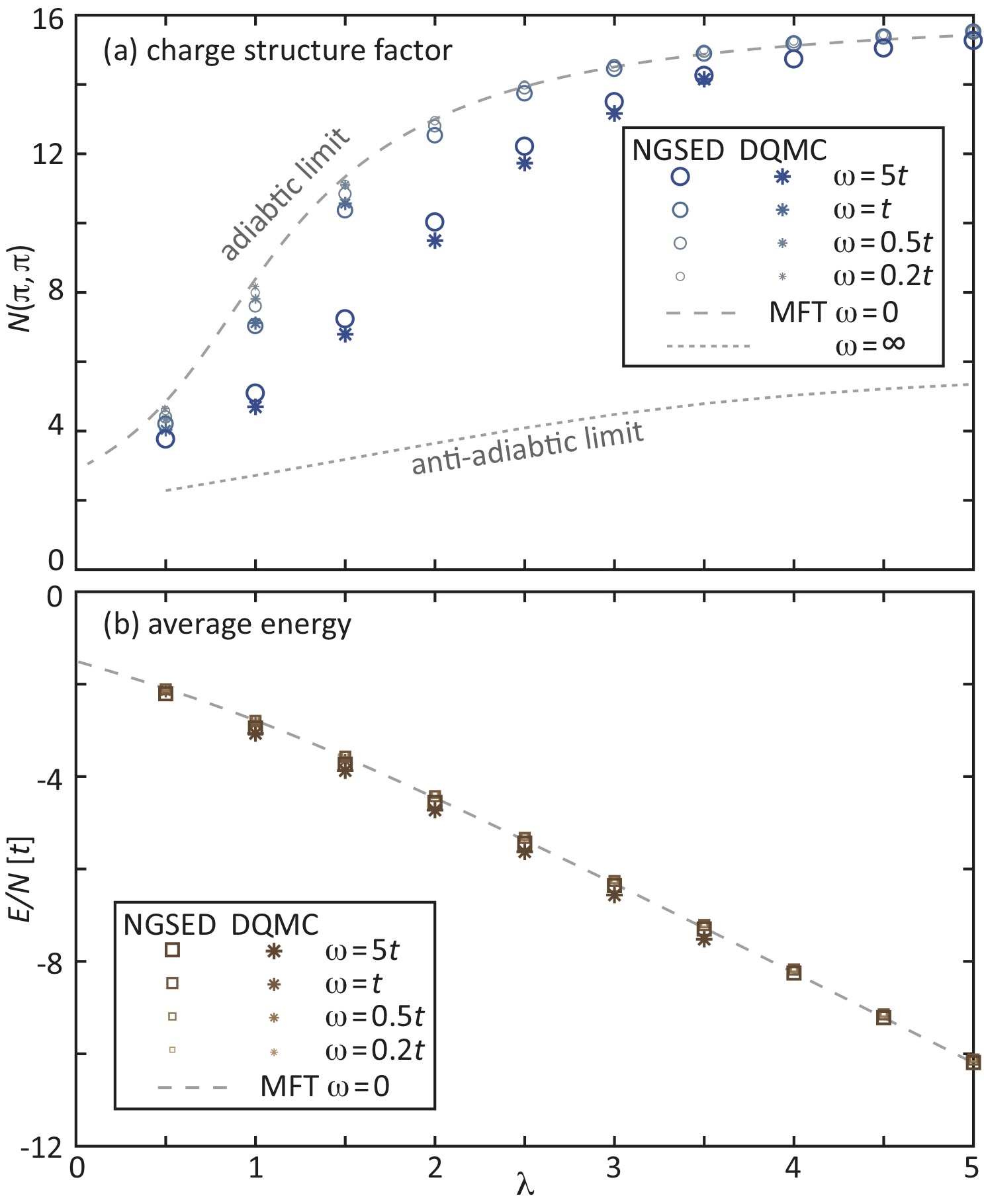}
\caption{\label{fig:holstein} The ground-state (a) charge structure factor $N(\pi,\pi)$ and (b) average energy per site $E/N$ as function of $\lambda$ in the Holstein model ($u=0$). The open dots are obtained by NGSED iterations, while the stars are obtained by DQMC with temperature extrapolated to $T=0$. The gray dashed lines indicate the adiabatic limit $\omega=0$ results obtained by MFT, while the gray dotted line represents the charge structure factor $N(\pi,\pi)$ for the anti-adiabatic limit $\omega=\infty$ obtained by the attractive Hubbard model.  }
\end{center}
\end{figure}
To make sure the wavefunction in Eq.~\eqref{eq:wvfuncansatzFull} correctly captures the phonon coupling in the e-ph system and does not induce significant bias on the electronic structure, we first benchmark our NGSED method with the pure Holstein model, \emph{i.e.}~for $u=0$, because it is a case where DQMC can give exact ground-state solutions (with extrapolation to $T=0$). Technically, DQMC is an unbiased numerical method for correlated fermionic models and is most efficient at high temperatures. The evaluation of low-temperature properties is usually bottlenecked by the fermion-sign problem, where the Boltzmann weight is not positive-definite. As an NP-hard (non-deterministic polynomial-time hard) problem, there are only a few models where the fermion-sign issue can be evadable, and the Holstein model is one example. Additional details regarding DQMC for the Holstein model are included in the Appendix~\ref{app:dqmcDetails}.

We compare the ground-state results for four different phonon frequencies -- $\omega=5t$, $t$, $0.5t$ and $0.2t$ -- obtained from NGSED and DQMC for the same temperature ($T=0$) and lattice size ($4\times4$). As shown in Fig.~\ref{fig:holstein}, the charge structure factor is monotonically increasing. With large $\lambda$s, the charge susceptibility approach $N=16$, which is the theoretical maximal value one can reach on a $4\times4$ cluster. In the thermodynamic limit, this charge susceptibility should always diverge with the presence of long-range charge order. For those parameter regimes accessible by DQMC, both the charge structure factor $N(\qbf)$ and the average energy $E/N$ match well between these two methods. For small $\omega$, DQMC becomes challenging at strong couplings due to prohibitively long autocorrelation times. Therefore, we compare the NGSED results with the mean-field theory (MFT) predictions for $\omega=0$, where MFT becomes exact. [see Appendix~\ref{app:adiabaticMFT} for the derivations]. We find the small-$\omega$ results asymptotically approach the MFT adiabatic predictions. Interestingly, the ground-state energy, with both electrons and phonons considered, is almost independent of the phonon frequency.

Another limit of the Holstein model is the anti-adiabatic limit, where the phonon frequency $\omega =\infty$. In this limit, the phonon degrees of freedom can be integrated out, leading to an instantaneous attraction between electrons. Unlike the phonon-mediated electronic interaction $V_\qbf$ in Eq.~\eqref{eq:effInteraction}, the attraction in the anti-adiabatic limit is $V=2\lambda$, independent of momentum $\qbf$\,\cite{hirsch1986enhanced,hirsch1986enhanced}. Therefore, it leads to an on-site attraction in real space. Due to the infinite phonon frequency, the dressing effect becomes a virtual process, indicating that the dressing correction to the kinetic energy vanishes. Therefore, in the $\omega=\infty$ limit, the problem exactly maps to the attractive Hubbard model with $U=-2\lambda$. It is where the Lang-Firsov transformation can exactly decouple the e-ph system. Since the phonon frequencies are all smaller or comparable to the electron bandwidth ($W=8t$) and we explicitly evaluate the phonon dressing effects, the ground state properties are far away from the anti-adiabatic limit. Comparison with the anti-adiabatic limit provides intuition for the ordering tendencies and will be further discussed in the context of the Hubbard-Holstein model.

\begin{figure}[!th]
\begin{center}
\includegraphics[width=0.9\columnwidth]{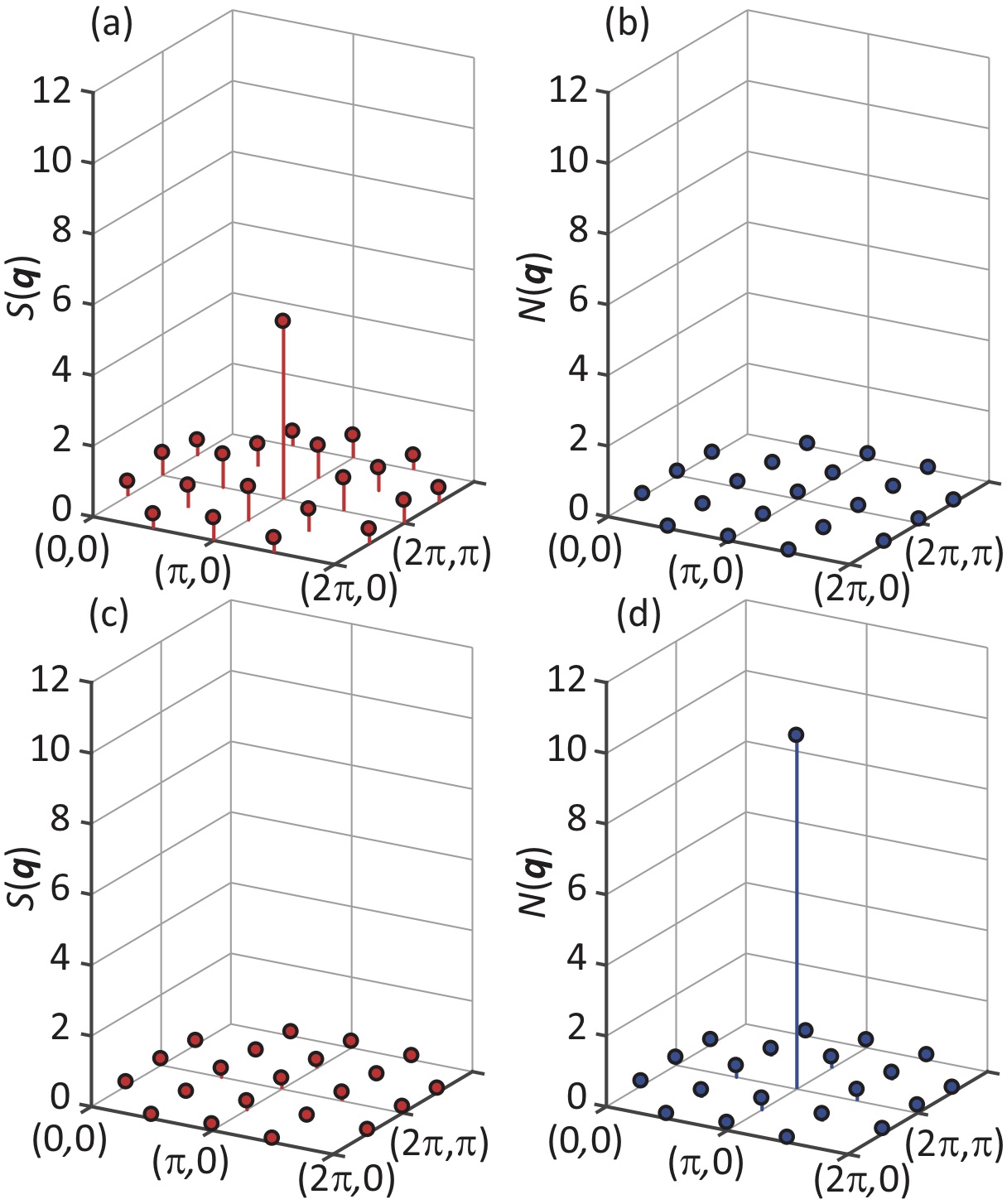}
\caption{\label{fig:nqsqDistr} (a,b) Distribution of (a) $S(\mathbf{q})$ and (b) $N(\mathbf{q})$ associated with the ground state obtained for $u=10$, $\lambda=4$ and $\omega=5$. (c,d) Same as (a,b) but for $u=0$. The $S(0,0)\equiv0$ and $N(0,0)\equiv16$ are not shown in the figure.
}
\end{center}
\end{figure}

The benchmarks with exact solutions obtained from DQMC and extreme limits in the Holstein model demonstrate that the NGSED method can adequately evaluate the coupling to phonons, though both the non-Gaussian transformation and phonon states are restricted to a variational subspace of the entire Hilbert space. The full wavefunction ansatz Eq.~\eqref{eq:wvfuncansatzFull} does not produce significant bias.

\begin{figure*}[!t]
\begin{center}
\includegraphics[width=2\columnwidth]{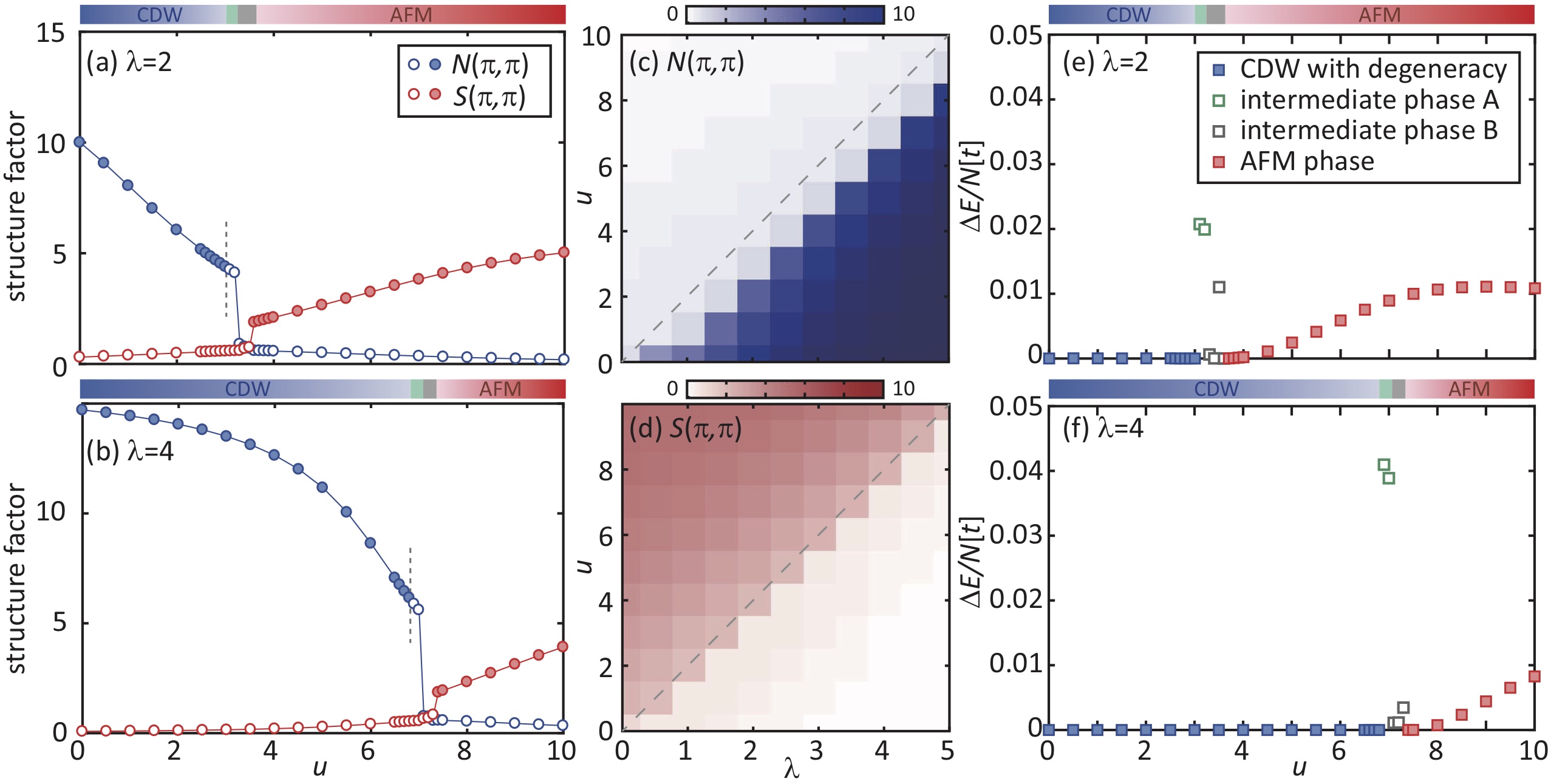}
\caption{\label{fig:phaseDiag} The ground-state $N(\pi,\pi)$ and $S(\pi,\pi)$ calculated for various $u$ values with (a) $\lambda=2$ and (b) $\lambda=4$. We distinguish the $S(\pi,\pi)$ data points for AFM and non-AFM phases as solid and open red circles. The solid (open) blue circles denote the $N(\pi,\pi)$ data points for those systems with (without) ground-state degeneracy, with the dashed line indicating the transition. 
(c,d) Diagram of (c) $N(\pi,\pi)$ and (c) $S(\pi,\pi)$ as a function of both $u$ and $\lambda$. The dashed lines denote the anti-adiabatic critical line $u=2\lambda$. 
(e,f) Energy gap per site $\Delta E/N$ as a function of $u$ for (e) $\lambda=2$ and (f) $\lambda=4$ same as (a,b): the open green and gray squares denote the two intermediate states (see classifications in the text). 
The upper bars for (a,b,e,f) guide the eye for these four regimes.
The phonon frequency $\omega$ is set as 5$t$ for all panels.
}
\end{center}
\end{figure*}

\subsection{Phase Diagram of the Hubbard-Holstein Model}
Having confirmed the accuracy of the method in the $u=0$ limit, we move on to finite $u$ and discuss the ground-state properties of the Hubbard-Holstein model. Although the phonon frequencies in typical correlated materials like cuprates are usually much smaller than the electronic bandwidth, here we first focus on a relatively high phonon frequency $\omega=5t$ for the purpose of elucidating the nature of the Hubbard-Holstein model. As indicated in Fig.~\ref{fig:holstein}, this selected frequency is away from both the adiabatic and anti-adiabatic limits. We will then discuss the frequency dependence in Sec.~\ref{sec:freqAndDoping}, where we will show that the interesting intermediate phases are less evident for smaller phonon frequencies.

A brief overview of the iteration processes for a few $\omega=5t$ systems are shown in Fig.~\ref{fig:evolution} of Sec.~\ref{sec:method_NGSiterations}: the ground state converges to two different phases for $u\gg\lambda$ and $u\ll\lambda$, as is seen from the structure factors. Here we present the detailed properties of this system and different phases.

Let us first look at these extreme cases, leaving the phases at $u\sim 2\lambda$ for later discussions. With the same set of parameter as Fig.~\ref{fig:evolution} [$\lambda=4$ and $\omega=5$], the momentum distribution of the ground-state spin and charge structure factors are shown in Fig.~\ref{fig:nqsqDistr}. For the $u$-dominant regime [here $u=10\gg \lambda$ for Figs.~\ref{fig:nqsqDistr}(a,b)], the system is dominant by the spin ordering, reflected by the large $S(\mathbf{q})$ compared with $N(\mathbf{q})$. More specifically, the spin correlation sharply peaks at $\mathbf{q}=(\pi,\pi)$ momentum. This reflects the tendency toward antiferromagnetism in the thermodynamic limit. At the same time, the system displays almost no charge fluctuations since the charge degrees of freedom are frozen at equilibrium. In the context of this paper, we call this spin-dominant phase an ``AFM phase'' though we do not have spontaneous SU(2) symmetry breaking in a finite cluster. In pure variational methods with mean-field decoupling, this AFM phase indeed establishes a symmetry breaking and a spin order parameter\,\cite{karakuzu2017superconductivity, ohgoe2017competition}. On the contrary, in the $\lambda$-dominant regime [here $u=0\ll \lambda$ for Figs.~\ref{fig:nqsqDistr}(a,b)], the ground state exhibits significant charge correlations. Different from the AFM case, here only the $(\pi,\pi)$ momentum exhibits strong correlations while other momenta are negligible. This is a difference between continuous and discrete symmetry breaking: the magnon fluctuations weaken the spin ordering in the AFM phase, while there is no Goldstone mode for the CDW phase. As expected, when charge ordering dominates, the ground state forms checkerboard doublons and holons, exhibiting no net spin correlation.

\begin{figure}[!th]
\begin{center}
\includegraphics[width=\columnwidth]{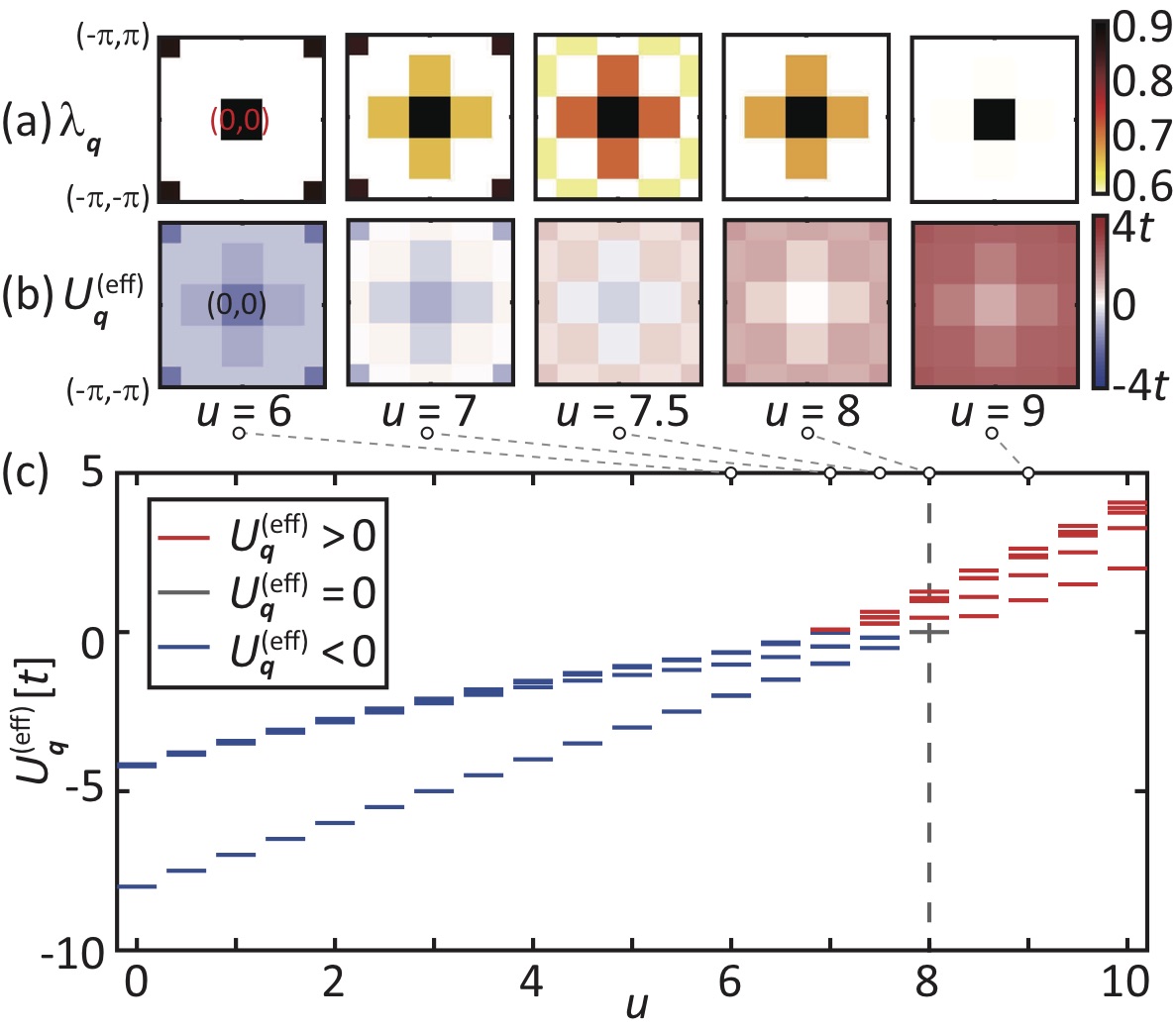}
\caption{\label{fig:vqDistrW5} (a,b) Distribution of the ground-state $\lambda_\qbf$ and $U^{\rm(eff)}_\qbf$ in the first Brillouin zone,  for $\lambda=4$, $\omega=5$ and $u=6$, 7, 7.5, 8 and 9. (c) Distribution of $U^{\rm(eff)}_\qbf$ for various $u$ values and $\lambda=4$, $\omega=5t$. The red bars represent $U^{\rm(eff)}_\qbf>0$, blue bars represent $U^{\rm(eff)}_\qbf<0$, and black bars represent $U^{\rm(eff)}_\qbf=0$. The lowest bar for each parameter set is always the $\qbf=0$ case. The dashed line indicates the anti-adiabatic limit $u=2\lambda$.
}
\end{center}
\end{figure}

With the increase of e-e interaction $u$ starting from the CDW phase for any fixed $\lambda$, the charge structure factor rapidly drops as shown in Figs.~\ref{fig:phaseDiag}(a) and (b). There is a sharp transition near $u\sim 2\lambda$ (but slightly away from this value, see discussions below). Beyond this transition point, spin correlations rapidly build up, overwhelm the charge instability, and form the Mott AFM state. For various $\lambda$'s and $u$'s, we obtain the coarse-grained ``phase diagram'' of spin and charge structure factors in Figs.~\ref{fig:phaseDiag}(c) and (d), indicating the regions of these two phases. Note that the difference between continuous and discrete symmetry breaking mentioned above leads to the distinct nature of the CDW and AFM phases. This is reflected by the ground-state degeneracy, or the excitation gap shown in Figs.~\ref{fig:phaseDiag}(e) and (f). In the CDW phase ($u\ll 2\lambda$), the ground state exhibits a two-fold degeneracy within the numerical accuracy; however, in the AFM phase ($u\gg 2\lambda$), the ground state is non-degenerate. This indicates, that the $4\times4$ system can be regarded as a $(\pi,\pi)$-ordered Peierls phase due to the commensurability and discrete symmetry breaking, while cannot support a SU(2) symmetry-breaking due to the power-law decay of spin correlations. 

The two extreme phases described above are expected and understood. What we are more interested in is the behavior near the boundary $u\sim 2\lambda$, where the two instabilities compete with each other. Interestingly, strong spin correlations start to build up already at $u<2\lambda$, as reflected in Figs.~\ref{fig:phaseDiag}(a) and (b). For example, in the $\lambda=4$ system, $S(\pi,\pi)$ becomes dominant at $u=7.4$ instead of 8; whereas in the $\lambda=2$ system, the AFM phase is reached for $u\geq 3.7$ instead of $4$. This is the case also for all $\lambda$s in the phase diagram in Figs.~\ref{fig:phaseDiag}(c) and (d). The fact that the boundary of the AFM phase sits on the $u<2\lambda$ side has been observed in 1D DMRG\,\cite{tezuka2007phase} and 2D QMC\,\cite{nowadnick2012competition} studies, but was not reproduced in previous variational studies with long-wavelength Lang-Firsov transformation. Now, with the NGSED, we are able to interpret the origin of this phenomenon explicitly. For convenience, let us define the effective Coulomb interaction as
\begin{equation}
    U^{\rm(eff)}_\qbf = U-V_\qbf.
\end{equation} 
In the anti-adiabatic limit, or Lang-Firsov picture, the sign of $U^{\rm(eff)}_\qbf$ determines the local trend to form a doublon or spin-singlet. We find this local picture being approximately correct for systems far away from the phase boundary, as shown in Fig.~\ref{fig:vqDistrW5}(b): the $U^{\rm(eff)}_\qbf$ are all negative for $u=6$ while positive for $u=9$. Although momentum-space fluctuation exists already in these cases, the ground state is qualitatively determined by the sign of the effective interaction. The overall sign accounts for the CDW and AFM at two extremes discussed above.

\begin{figure}[!t]
\begin{center}
\includegraphics[width=\columnwidth]{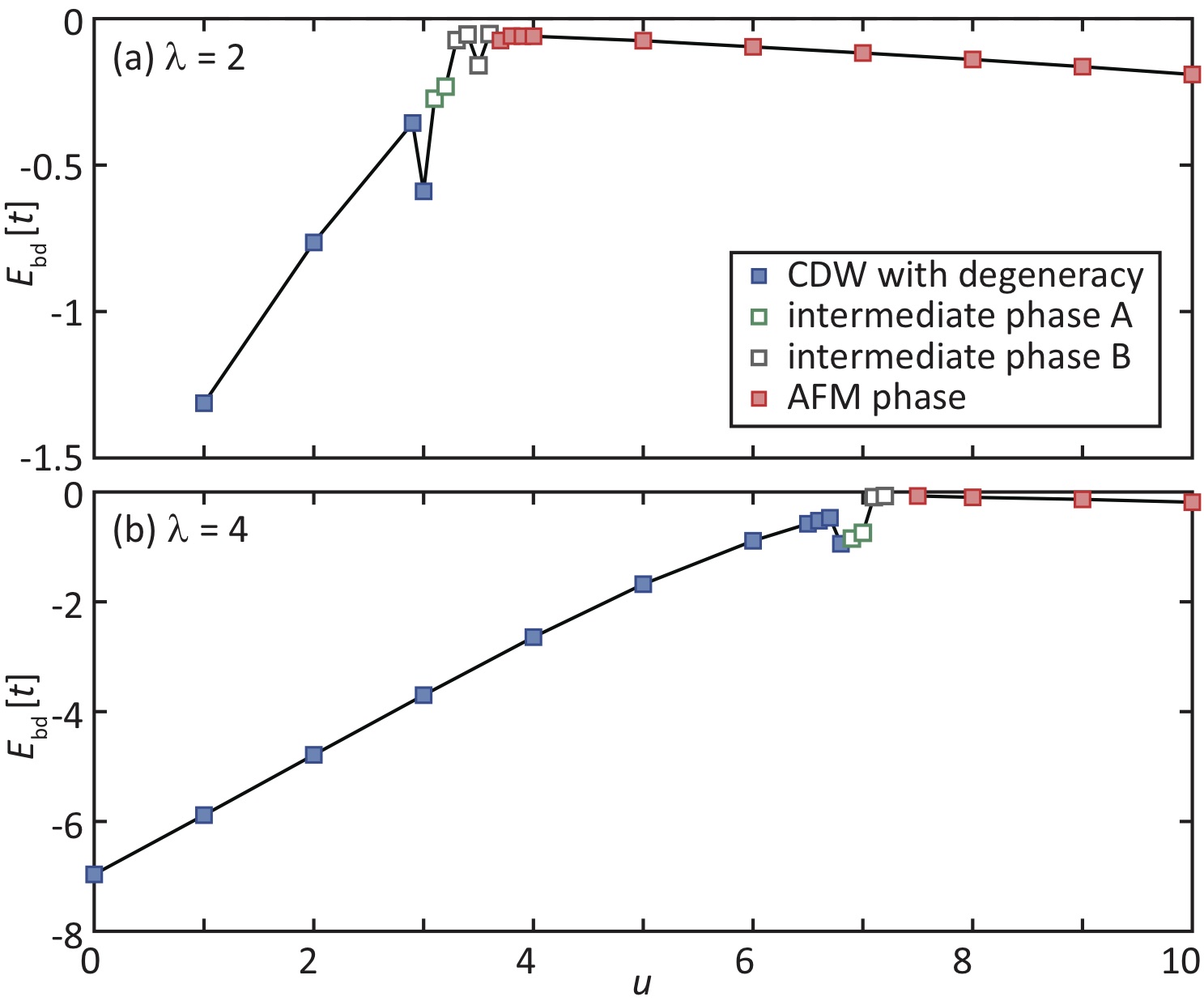}
\caption{\label{fig:bindingEnergy} The binding energy as a function of $u$ for (a) $\lambda=2$ and (b) $\lambda=4$ and $\omega=5$ [same parameter set as Figs.~\ref{fig:phaseDiag}(a,e) and \ref{fig:vqDistrW5}].
}
\end{center}
\end{figure}

However, the ground-state solution for the polaronic dressing parameter $\lambda_\qbf$ strongly varies over the first Brillouin zone [see Fig.~\ref{fig:vqDistrW5}(a)]. Due to the large charge susceptibility, the polaronic dressing, reflected in $\lambda_\qbf$, converges to a substantially larger value at the nesting momentum than other $\qbf$s. In contrast to the uniform distribution assumed in the Lang-Firsov transformation and the $U^{\rm(eff)}_\qbf\equiv U-2\lambda$ consequence, such momentum fluctuations of $\lambda_\qbf$ leads to the effective long-range interactions $V_\qbf$ and, accordingly, the fluctuations of $U^{\rm(eff)}_\qbf$. These momentum fluctuations may not be critical when $U^{\rm(eff)}_\qbf$ is significantly positive or negative (\emph{i.e.}~for $u\ll2\lambda$ or $u\gg2\lambda$), but plays a role near the boundary between CDW and AFM phases.
As shown in Figs.~\ref{fig:vqDistrW5}(b) and (c), the $U^{\rm(eff)}_{\qbf=(0,0)}$ is always lower than other momenta [at the CDW phase $U^{\rm(eff)}_{\qbf=(\pi,\pi)}\! \approx\! U^{\rm(eff)}_{\qbf=(0,0)}$]. Moreover, only the effective repulsion at $\qbf=(0,0)$ follows the anti-adiabatic prediction as a function of $u$, while other $U^{\rm(eff)}_\qbf$ are much larger. Therefore, the anti-adiabatic phase boundary $u=2\lambda$, where the strength of phonon-induced interaction is estimated by the local Lang-Firsov transformation, overestimates the realistic impact of phonons. The consequence is that charge ordering drops and spin ordering develops at a relatively small $u$ value.

Apart from the shift of the phase boundary, the fluctuations of the effective interactions lead to exotic intermediate phases near the phase boundary. As indicated in Fig.~\ref{fig:phaseDiag}, there are two narrow regimes between the well-determined CDW and AFM phases. One intermediate regime (denoted as A) lies next to the CDW phase (\emph{e.g.},~$3.1\leq u\leq 3.2$ for $\lambda=2$ and $6.9\leq u\leq7$ for $\lambda=4$), marked as green in Fig.~\ref{fig:phaseDiag}. In this regime, the system still exhibits a large charge structure factor, but has lost the ground-state degeneracy, \emph{e.g.}, displays finite excitation gap. That being said, the system lies in a non-CDW state with large charge structure factor in this narrow regime. From the perspective of $U_\qbf^{\rm(eff)}$, it is because $U_\qbf^{\rm(eff)}$ has changed sign in part of the Brillouin zone, though it is still negative at the nesting momentum.

The situation in the intermediate regime A is very similar to the Luther-Emery liquid in 1D or quasi-1D system, which might display coexisting superconductivity and charge order\,\cite{Greitemann2015arxiv, jiang2018superconductivity}. In recent VMC studies, the entire intermediate regime was claimed to be superconducting\,\cite{karakuzu2017superconductivity, ohgoe2017competition}. However, due to the biased electronic wavefunction ansatz, the conclusion remains controversial\,\cite{hohenadler2019dominant}. With the full many-body wavefunction kept for the electrons, the NGSED calculation provides a more reliable characterization of the two intermediate regimes. Unfortunately, we cannot examine the scaling of the charge or pair correlations and extract the correlation length in a finite cluster. As a compromise, we calculate the binding energy defined as 
\begin{equation}
    E_{\rm bd} = E(\textrm{half-filling}) + E(2\textrm{-hole}) - 2E(1\textrm{-hole})\,.
\end{equation}
Figure~\ref{fig:bindingEnergy} shows the evolution of $E_{\rm bd}$ as a function of $u$ for $\lambda=2$ and $\lambda=4$. For $u<3$ ($\lambda=2$) and $u<6.8$ ($\lambda=4$), the binding energy is sizable and negative, {as the CDW state forms bipolarons with strong couplings. This binding energy} decreases (in magnitude) with the rise of $u$. {Up to intermediate regime A, the ground state is fragile due to the competition between two different instabilities, but we still observe finite binding energy.} This might be an indication of coexisting Cooper pairs. Intriguingly, recent DQMC studies suggested that the phonon dispersion in the Holstein model may favor superconductivity over CDW\,\cite{costa2018phonon}. Such a dispersive (effective) phonon energy is indeed realized in the Hubbard-Holstein model through the momentum dependence of $\lambda_\qbf$ [see Eq.~\eqref{eq:renormalizedPhononFreq}] and becomes more evident in the intermediate regimes, as discussed above. Therefore, both the binding energy and phonon dispersion indicate the possible presence of superconductivity in the intermediate regimes.

More precisely, in contrast to intermediate regime A, intermediate regime B, which is adjacent to the AFM phase, exhibits both small charge and spin structure factors. This phenomenon is distinct from the 1D Hubbard-Holstein model. Here, in the intermediate regime B, the effective interactions have been delicately balanced and become too weak to overcome the kinetic energy and localize electrons. Considering the possible tendency toward superconductivity present in the neighboring intermediate regime A, regime B is possibly superconducting.

\begin{figure}[!t]
\begin{center}
\includegraphics[width=0.95\columnwidth]{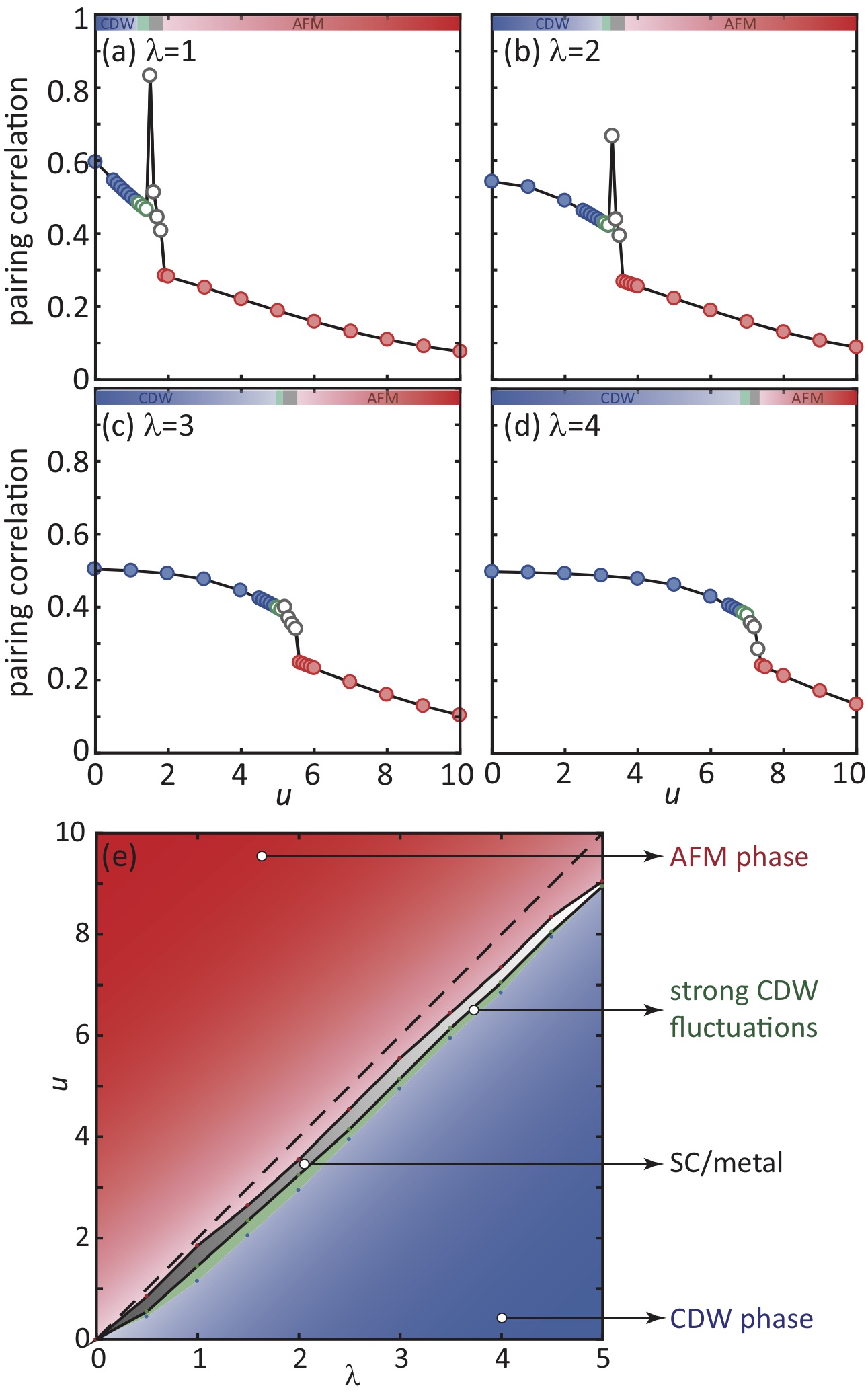}
\caption{\label{fig:DeltaSPairing} The pairing correlation calculated for half-filled Hubbard-Holstein model with (a) $\lambda=1$, (b) $\lambda=2$, (c) $\lambda=3$, and (d) $\lambda=4$ for various $u$s and $\omega=5t$. (e) The phase diagram of the 2D Hubbard-Holstein model. The red and blue regimes denote the AFM and CDW phases, respectively; the green regime (intermediate regime A) represents the non-degenerate regime with strong CDW fluctuations, while the gray regime (intermediate regime B) is superconducting or metallic. The darkness of colors guides the eye for the strength of the corresponding orders. The gray dashed line denotes the anti-adiabatic phase boundary $u=2\lambda$.
}
\end{center}
\end{figure}

However, the binding energy discussed above should be rigorously defined in the thermodynamic limit, where the addition of carriers can be treated as a perturbation. (In the 4$\times$4 cluster, each doped hole represents a 6.25\% doping.) Therefore, we rely on the pairing correlations instead of $E_{\rm bd}$ in the determination of the phases in the half-filled system. To investigate the nature of the intermediate regimes and conduct a more rigorous analysis of superconductivity, we calculate the $s$-wave superconducting pair correlation function, defined as 
\begin{eqnarray}
   P_s = \frac1N \big\langle \Psi \big| \Delta_s^\dagger \Delta_s\big|\Psi\big\rangle\,,
\end{eqnarray}
where the pairing operator is
\begin{eqnarray}
  \Delta_s = \sum_\ibf   c_{\ibf\downarrow} c_{\ibf\uparrow} = \sum_\kbf c_{-\kbf\downarrow} c_{\kbf\uparrow}\,.
\end{eqnarray}
Note, the expectation value should be taken over the full wavefunction $|\Psi\rangle$ instead of just the electronic wavefunction $|\psi_e\rangle$. Different from the charge and spin structure factor, the pairing operator $\Delta_s$ does not commute with the non-Gaussian transformation $\Upl$. Thus, the expansion of the pairing correlation function, with electron wavefunction and variational parameters, is
\begin{eqnarray}\label{eq:pairingcorr}
	 P_s\! &=& \!\frac1{N}\!\sum_{\ibf,\jbf} \langle\psi_e| c_{\ibf\uparrow}^\dagger  c_{\ibf\downarrow}^\dagger   c_{\jbf\downarrow} c_{\jbf\uparrow} |\psi_e\rangle 
	 e^{\sum_\qbf\!\frac{4|\lambda_\qbf|^2}{N} e_2^T \!\Gamma_\qbf\! e_2 (e^{i\qbf(\rbf_\ibf\!-\!\rbf_\jbf)}-1)}\nonumber\\
	 &=& \!\frac1{N^2}\sum_{\Qbf,\kbf,\kbf^\prime} \langle\psi_e|  c_{\kbf^\prime\uparrow}^\dagger  c_{\Qbf-\kbf^\prime\downarrow}^\dagger   c_{\Qbf-\kbf,\downarrow} c_{\kbf,\uparrow} |\psi_e\rangle\nonumber\\
	 && \times\sum_{\rbf}e^{-i\Qbf\cdot\rbf}
	 e^{\sum_\qbf\frac{4|\lambda_\qbf|^2}{N} e_2^T \Gamma_\qbf e_2 (e^{i\qbf\cdot\rbf}-1)}\,.
\end{eqnarray}
While permuting the electronic operators with the polaronic transformation $\Upl$, they physically represent the same operators of dressed quasiparticles. Therefore, to evaluate the BCS-type electronic pairs, one has to compute a superposition of FFLO-type quasiparticle pairs, because the polaronic dressing exchanges momentum between electrons and phonons.

Figure~\ref{fig:DeltaSPairing} presents $P_s$ as a function of $u$ calculated for four different $\lambda$'s. The pairing correlation is close to 0.5 for $u=0$, which is the expectation value for a CDW state; it is strongly suppressed in the AFM phase due to the low rate of double occupancy. We observe an enhancement of $P_s$ in the intermediate regime B. {As shown in Appendix~\ref{app:superconductivity}, the enhancement is only evident for $s$-symmetry.} This enhancement is relatively large for small $\lambda$, supporting the existence of superconductivity. With the increase of coupling strength, both the pairing correlation {and the coherence of Cooper pairs} in the intermediate regime A are gradually suppressed until $\lambda\sim3$, where it becomes a smooth crossover between the CDW and AFM phases (up to the parameter resolution selected in our calculation $\Delta u=0.1$). If the intermediate regime B is indeed superconducting, the regime A with strong CDW fluctuations could be a crossover between CDW and superconductivity.

To summarize the evolution of superconductivity and the intermediate regimes, we sketch a phase diagram in Fig.~\ref{fig:DeltaSPairing}(e) through a grid of $\Delta u=0.1$ and $\Delta \lambda=0.5$. The two intermediate regimes are denoted as green (intermediate regime A) and gray (intermediate regime B), following the same color code as Fig.~\ref{fig:phaseDiag}. We exploit the variation of darkness to represent the change of pairing correlations in the intermediate regime B. In our calculations we find the intermediate regimes are enlarged only slightly as $\lambda$ decreases from 5 to 0.5. This is in contrast with the VMC predictions which assigned the entire $u<2\lambda$, $\lambda<1$ region as superconducting\,\cite{karakuzu2017superconductivity, ohgoe2017competition}. However, our results are consistent with the QMC conclusions in the same regime\,\cite{hohenadler2019dominant}. Considering that QMC is unbiased, this conclusion reflects the necessity of reliability treating the electronic wavefunction. More recently, our phase diagram in Fig.~\ref{fig:phaseDiag} has been confirmed by an independent QMC study with finite-size scaling\,\cite{costa2020phase}, suggesting that the intermediate state indeed exhibits long-range ordered superconductivity. This consistency further demonstrates the reliability of our NGSED method and the phase assignment using a finite cluster.

\subsection{Impact of Phonon Frequencies and Doping}\label{sec:freqAndDoping}
\begin{figure}[!t]
\begin{center}
\includegraphics[width=\columnwidth]{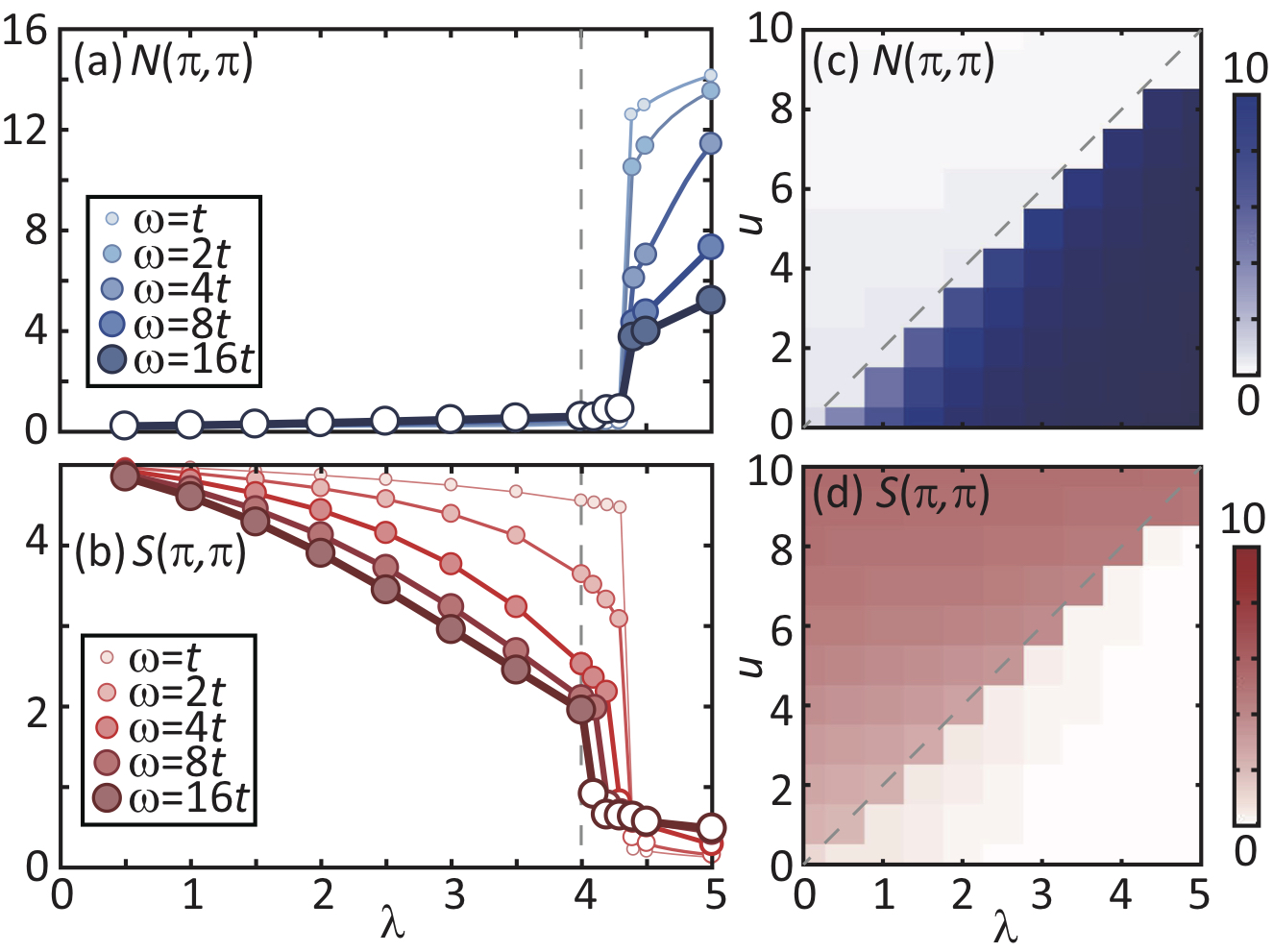}
\caption{\label{fig:phaseVariousWU} (a) Charge and (b) spin structure factor as a function of dimensionless e-ph coupling $\lambda$ for $u=8$ and $\omega=t$, 2$t$, 4$t$, 8$t$, and 16$t$, respectively. Diagram of (c) $N(\pi,\pi)$ and (d) $S(\pi,\pi)$ as a function of both $u$ and $\lambda$. The dashed lines denote the anti-adiabatic critical line $u=2\lambda$. The phonon frequency $\omega$ is set as $t$.
}
\end{center}
\end{figure}

Having understood the phase diagram of the 2D Hubbard-Holstein model with a fixed phonon frequency, we briefly discuss the impact of various frequencies and carrier doping in this subsection.

Similar to the case of the Holstein model discussed in Sec.~\ref{sec:Holstein}, we expect the e-ph system exhibiting steeper phase transitions with smaller phonon frequency. For a fixed $\lambda$, the smaller $\omega$ implies larger $g/\omega$. As shown in Figs.~\ref{fig:phaseVariousWU}(a) and (b) for the calculations with $\omega$ ranging from $t$ to 16$t$, both charge and spin structure factors drop more rapidly for smaller frequencies when approaching the phase boundary, consistent with previous DQMC results at finite temperatures\,\cite{nowadnick2012competition, johnston2013determinant}. Intuitively, it can be understood as the adiabatic limit behaves similar to a mean-field theory, suppressing all quantum fluctuations which accumulates before reaching a phase transition. Here, using the language of the polaronic dressing in the non-Gaussian wavefunction, we provide the interpretation from a different perspective -- the combined impact of polaronic dressing in both tunneling and interaction parameters. As is well known, the Lang-Firsov transformation should give the same effective e-e interaction for a fixed $\lambda$ in the atomic limit. However, the dressing parameter $\lambda_\qbf$, to generate the same $V_\qbf$, is larger for a smaller $\omega$. That means, if one takes the tunneling terms into account, the polaronic renormalization for $t_\alpha$ is larger. Therefore, the quantum fluctuations become effectively weaker with respect to the same interaction strength, leading to a sharper phase transition.

For the same reason, for larger phonon frequencies, the boundary of the AFM phase is less affected by spatial fluctuation of $V_\qbf$, therefore, it is closer to the $u=2\lambda$ anti-adiabatic line. Such a move of the phase boundary causes a larger intermediate phase, consistent with the VMC results\,\cite{karakuzu2017superconductivity, ohgoe2017competition}. That being said, the intermediate phase becomes invisible (if it exists at all) for smaller phonon frequencies comparable to those in cuprates. Limited by the resolution and finite size of our calculation, we are unable to determine whether a critical coupling strength exists for smaller phonon frequencies.

\begin{figure}[!b]
\begin{center}
\includegraphics[width=\columnwidth]{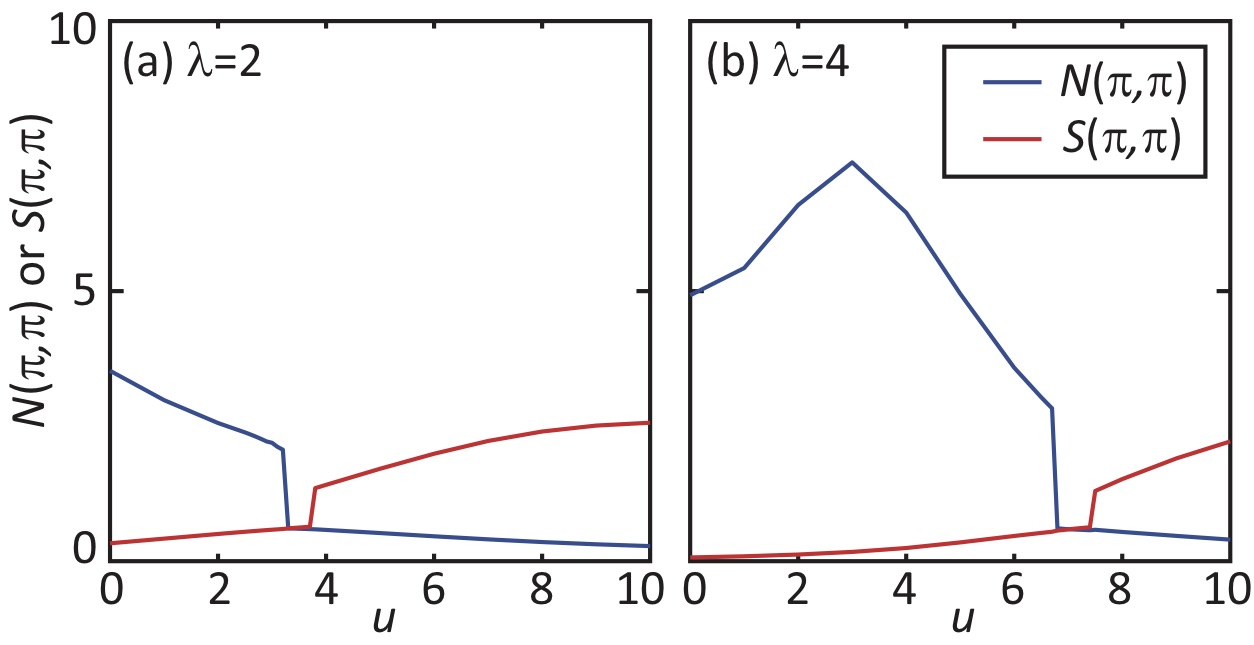}
\caption{\label{fig:doping} Structure factors $N(\pi,\pi)$ and $S(\pi,\pi)$ calculated for 12.5\% doping: for various interaction parameters $u$ and (a) $\lambda=2$, (b) $\lambda=4$. The phonon frequency is $\omega=5t$. 
}
\end{center}
\end{figure}

Varying both $\lambda$ and $u$ for $\omega=t$, we obtain the phase diagram shown in Figs.~\ref{fig:phaseVariousWU}(a) and (b) using NGSED. An immediate observation is the suppression of the intermediate regime, if it exists at all. This regime is invisible in the VMC studies on the strong-coupling side\,\cite{karakuzu2017superconductivity,ohgoe2017competition}, but is still present at finite temperature according to DQMC studies\,\cite{nowadnick2012competition}. Although the effective interaction is more dispersive [see Appendix~\ref{app:smallPhonon}], its impact on the electronic configuration becomes less critical, due to the suppression of quantum fluctuations as mentioned above. This accounts for the similarity of the phase diagram compared with previous ED calculations on a Peierls-Hubbard model [with only $\qbf=(\pi,\pi)$ phonon mode]\,\cite{wang2018light}. It is worth mentioning that the convergence for smaller $\omega$ requires many more iterations since the lack of quantum fluctuations causes traps in local energy minima in the parameter space. The convergence speed can be improved by a few warm-up iterations, as discussed in Appendix~\ref{app:smallPhonon}.

With the presence of finite doping, the competition between spin and charge order is not restricted to a single nesting momentum. Although both $N(\qbf)$ and $S(\qbf)$ spread out in momentum, the $\qbf =(\pi,\pi)$ component still dominates [the structure factors calculated at other $\qbf$'s are all smaller than 1.5, not shown here].  Figure~\ref{fig:doping} shows calculations for 12.5\% doping with $\lambda=2$ and 4. Both the charge and spin structure factors are significantly smaller than the half-filled case [see Fig.~\ref{fig:phaseDiag}]. Interestingly, the ground-state charge structure factor for $\lambda=4$ is not monotonically suppressed by the increase of $u$, in contrast to the situation at half-filling. For $u<4$, the increase of electron correlations in fact slightly enhances the $(\pi,\pi)$-charge ordering. This trend may be regarded as a correlation-enhanced polaronic dressing effect\,\cite{mishchenko2004electron}: the presence of electronic correlations reduces the mobility of carriers in a doped system, and therefore favors the polaronic dressing to some extent. A more rigorous confirmation of this non-monotonicity and a specific assessment of the underlying physics are beyond the scope of this work, and should be further investigated using a combination of multiple numerical methods.

\section{Conclusion and Outlook}\label{sec:conclusion}
We have presented NGSED, a wavefunction-based method used to treat systems with both e-e and e-ph interactions, taking advantage of both variational non-Gaussian transformations and exact diagonalization. The variational part of the wavefunction avoids the challenge of treating an excessively large Hilbert space for phonons, while the full many-body electronic state minimizes bias and allows for the complexities associate with electronic correlations. We presented the formalism for this method using a generic e-ph system, where the e-ph coupling is $g_\qbf$, the e-e interaction is $U_\qbf$ and the phonon energy is $\omega_\qbf$ are allowed to be momentum dependent. We applied the NGSED method to the Hubbard-Holstein model, where we compare with various other approaches. To assess the bias incurred by our variational ansatz we have benchmarked against numerically exact DQMC results on the Holstein model. The consistency with DQMC results justifies the correctness of NGSED, at least for the Holstein type of e-ph coupling.

With this new method, we have examined the ground-state properties of the 2D Hubbard-Holstein model. While in the limiting cases where one of the interactions is dominant, our results are consistent with known conclusions, we have found interesting and delicate structures near the transition. We show that the boundary of the AFM phase is on the $u<2\lambda$ side, which is consistent with the known exact results in 1D and variational results in 2D, but has not been completely explained yet. With the information of the entangled e-ph wavefunction, we provided an intuitive picture of this boundary shift from the effective e-e interaction point of view. We demonstrate that the traditional local Lang-Firsov transformation overestimates the impact of phonons by neglecting their uneven momentum distribution. The advantage of the NGSED method is its efficacy for capturing this distribution and physically addressing the origin of the boundary shift. 

In addition to the boundary shift, we have identified two narrow intermediate regimes between the CDW and AFM phases. One of them may be superconductivity, while the other exhibits strong charge fluctuations and significant binding energy. Both phases reside within the superconducting phase suggested by VMC studies\,\cite{karakuzu2017superconductivity, ohgoe2017competition}. However, the intermediate regimes obtained in our NGSED calculations are much narrower and do not intersect with $u=0$, a result that is supported by unbiased QMC calculations. Although the 2D Hubbard-Holstein model is the simplest toy model involving both electron-electron and electron-phonon interactions, the presence of enhanced superconductivity in the intermediate regime may be related to the superconducting dome in high-$T_c$ cuprates, as recent observations have indicated the important role of phonons in the overdoped regime\,\cite{he2018rapid}.

With the capability to adequately capture the phonon dressing, the NGSED method combines the merits of variational and exact approaches in many-body systems: it addresses the issue of both the large phonon Hilbert space and the lack of correlations in the pure variational approach. Thus, it provides a general prototype for a variety of problems involving e-e and e-ph interactions: by allowing the coupling strength $g_\qbf$ and phonon energy $\omega_\qbf$ to vary in momentum space, it can be applied to more realistic e-ph systems like those with forward scattering, $B_{1g}$ or acoustic phonons; by a rotation of the fermionic basis via the $\Upl$ similar to Eq.~\eqref{eq:pairingcorr}, it can also be employed to calculate other instantaneous observables involving high-order correlations. More importantly, as a wavefunction-based method, the NGSED method can be generalized to investigate the out-of-equilibrium physics in the pump-probe electron-phonon system, through the projection of equations of motion for the real-time dynamics \,\cite{shi2019ultrafast} combined with advanced Krylov-subspace techniques. By the same means, it can also be extended to the evaluation of excited states, spectroscopies, and thermal ensembles.\,\cite{guaita2019gaussian,shi2019variational, hackl2020geometry} With these extensions, NGSED may be used to explain and predict complex spectroscopies and pump-probe experiments, which are beyond the capability of perturbation and statistical methods.

The polaron transformation provides the lowest order decoupling between electrons and bosons. Extending to more intricate forms of non-Gaussian transformations, the NGSED method can be employed to decouple the interaction between electrons and other bosonic excitations, such as excitons, plasmons, and magnons. The non-Gaussian transformations have been used to study impurity models like the Kondo and Anderson models\,\cite{ashida2018variational,ashida2018solving,shi2019ultrafast}, and some models in lattice gauge theory, like the 1D Schwinger model\,\cite{sala2018variational}, paving the way for application to Kondo-Hubbard and Anderson-Hubbard models, as well as the lattice gauge theory in higher dimensions. The study of these electron-boson or impurity problems would help to elucidate the collective and local properties of correlated materials.

More generally, numerical methods involving non-Gaussian wavefunctions offer opportunities to extend electronic structure theory. The traditional \textit{ab initio} electronic structure theory is constructed on top of Gaussian states (Slater determinants), evolving into post-Hartree-Fock methods (configuration interaction, coupled cluster, \textit{etc.}) and multi-reference methods. Using the non-Gaussian wavefunctions as the fundamental basis, one can embed quantum entanglement at the outset. The NGSED method, as an analog of the full configuration interaction, can be regarded as the first building block in a non-Gaussian-based post-mean-field class of methods. Relevant post-mean-field methods constructed on this set of bases include the embedding with other many-body approaches. For example, with the same formalism handling the phonon wavefunction, the non-Gaussian transformation can be embedded with DMRG or iPEPS\,\cite{corboz2016improved}, self-consistently transforming a fermion-boson problem into one of quasiparticles with long-range interactions. Since the non-Gaussian transformation has rotated the many-body basis from electrons to quasiparticles, it might be helpful to reduce the fermion-sign issue in DQMC. Moreover, the multi-reference framework can also be extended to a non-Gaussian wavefunction basis, through the construction of superpositions of non-Gaussian wavefunctions or even NGSED.

\section*{Acknowledgements}
We thank Y. Ashida, L. Hackl, and F. Liu for insightful discussions. Y.W. and E.D. acknowledge the National Science Foundation through Grants No.~DMR-2038011 and No. OAC-1934714, ARO Grant No.~W911NF-20-1-0163, and the Harvard-MIT Center for Ultracold Atoms. Y.W. acknowledges the Postdoctoral Fellowship in Quantum Science of the Harvard-MPQ Center for Quantum Optics. I.E. acknowledges support from the Harvard Quantum Initiative Postdoctoral Fellowship in Science and Engineering. J.I.C. is supported by the EU through the ERC Advanced Grant QUENOCOBA (Grant No.~742102). This research used resources of the National Energy Research Scientific Computing Center (NERSC), a U.S. Department of Energy Office of Science User Facility operated under Contract No. DE-AC02-05CH11231.

\bibliography{paper}

\begin{thebibliography}{80}
\bibitem [{gen()}]{genericEPC}%
  \BibitemOpen
  \href@noop {} {}\bibinfo {note} {The most generic form of e-ph interaction
  involves all quadratic electronic terms, not restricted to the density
  operator. Therefore, the generic e-ph coupling is $g_{\kbf\qbf}$ with
  electronic momentum dependence.}\BibitemShut {Stop}%
\bibitem [{zer()}]{zerothOrderTVec}%
  \BibitemOpen
  \href@noop {} {}\bibinfo {note} {The zeroth-order tangential vector $|0_{\rm
  ph}\rangle\otimes|\psi_{\rm e}\rangle$ has no projection on the right-hand
  side due to the cancellation of the average energy. This part reflects the
  contribution of pure electronic states and has been optimized in
  Sec.~\ref{sec:electronOptimization}.}\BibitemShut {Stop}%
\end{thebibliography}

\appendix
\setcounter{figure}{0}
\renewcommand\thefigure{A\arabic{figure}}
\section{Brief Introduction to DQMC in the Holstein model}\label{app:dqmcDetails}

\begin{figure}[!th]
\begin{center}
\includegraphics[width=\columnwidth]{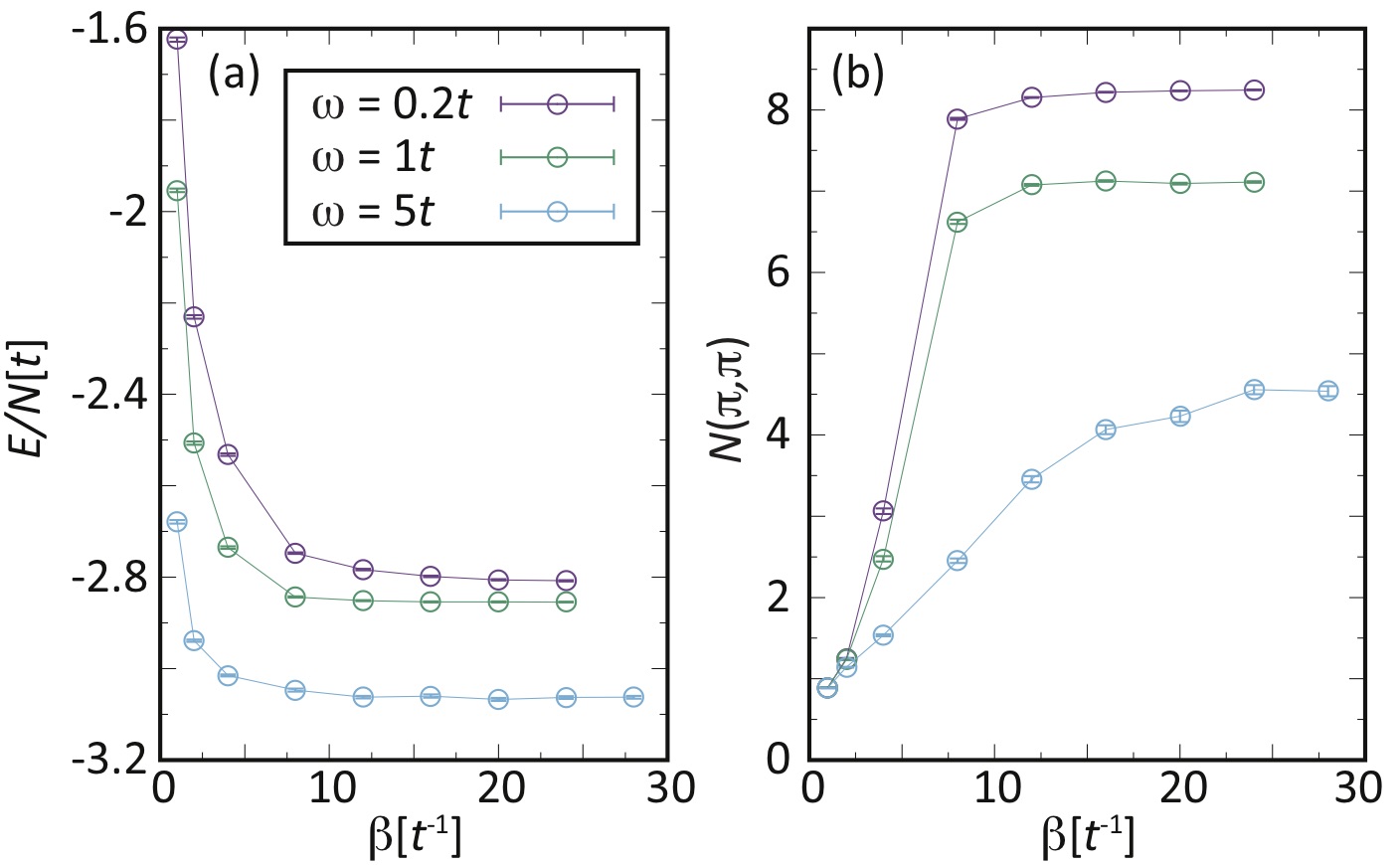}
\caption{\label{fig:T0_extrap} (a) Average energy $E/N$ and (b) structure factor $N(\pi,\pi)$ as a function of inverse temperature $\beta = 1/T$ for representative parameters $\lambda=1$ and $\omega=0.2,1,5$. In both plots errorbars are smaller than the symbol size. For phonon frequencies $\omega \lesssim t$ these observables attain their asymptotic $T=0$ value for $\beta \gtrsim 20$. For larger phonon frequencies lower temperatures are required. 
}
\end{center}
\end{figure}

We present a brief introduction and supplementary data about the DQMC technique. As a standard technique to many-body systems, especially the Holstein model, a detailed introduction to DQMC can be found in literature, \emph{e.g.}, Refs.~\onlinecite{PhysRevD.24.2278, PhysRevB.40.506, johnston2013determinant}. We emphasize that the notorious fermion sign problem is absent in the Holstein model because the phonon field couples in the same way to both spin-up and -down electrons, so that the probability measure is proportional to the fermion determinant squared and is therefore non-negative. The absence of a sign problem allows us to access relatively low temperatures. However, at exceedingly low temperatures DQMC calculations for the Holstein model are still limited due to prohibitively long autocorrelation times \,\cite{hohenadler2008autocorrelations}. To partially mitigate this issue, we employ a combination of local and global updates, as explained in Ref.~\onlinecite{johnston2013determinant}.

In Fig.~\ref{fig:T0_extrap} we show how the energy density $E/N$ and structure factor $N(\pi,\pi)$ approach their asymptotic $T=0$ values for representative parameters $\lambda = 1$ and $\omega = 0.2,1,5$. For all the DQMC data reported in the main text, we use the values of $E/N$ and $N(\pi,\pi)$ at our lowest temperature, where they have ceased to change appreciably, to approximate the value at $T=0$. We note that that the requisite temperatures to probe the $T=0$ limit become lower as $\omega$ increases. This trend can be understood from the fact that the Holstein model maps to the negative-$U$ Hubbard model in the limit $\omega \to \infty$, which has a vanishing $T_c$ for coexisting SC and CDW order. On the other hand, for $\omega=0$, $T_c$ is roughly on the order of the hopping $t$.

\setcounter{figure}{0}
\renewcommand\thefigure{B\arabic{figure}}
\section{Exact Mean-Field Solutions in the Adiabatic Limit}\label{app:adiabaticMFT}
In this appendix, we provide the derivation of the MFT solution in the adiabatic ($\omega=0$) limit.
\begin{figure}[!b]
\begin{center}
\includegraphics[width=\columnwidth]{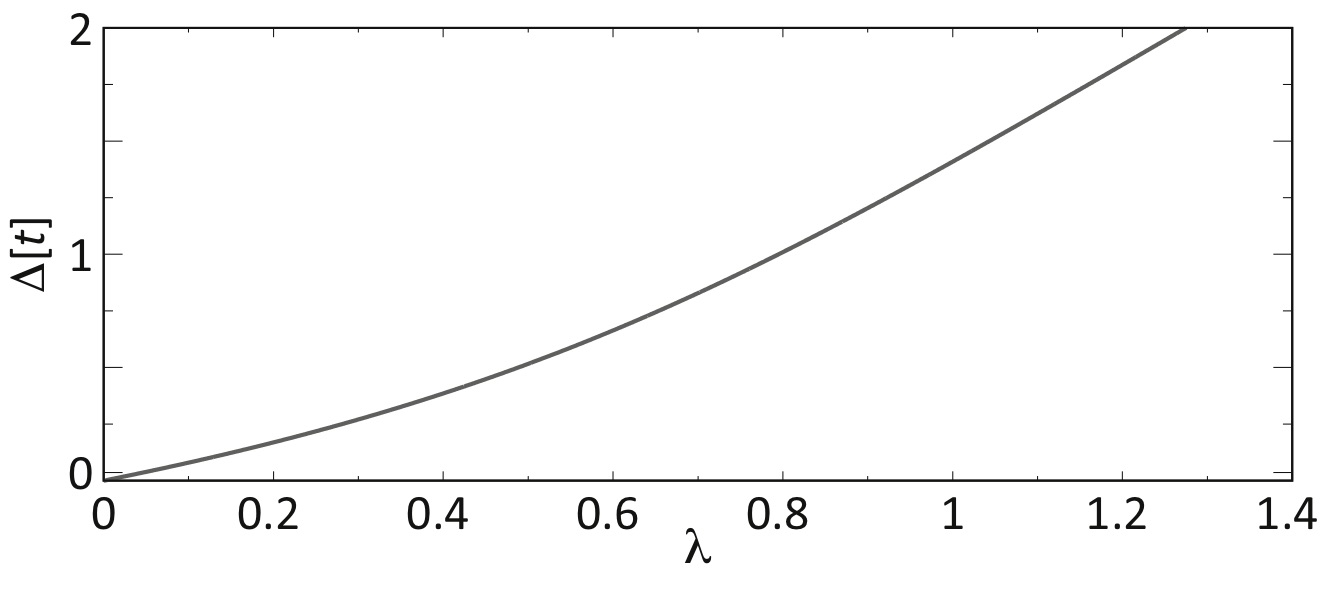}
\caption{\label{fig:del_vs_lam}  CDW gap $\Delta$ as a function of $\lambda$ in the adiabatic $\omega = 0$ limit, obtained by solving \eqref{eq:gap_eq}.
}
\end{center}
\end{figure}

In the adiabatic limit of infinite ion mass, corresponding to $\omega\to 0$, mean-field theory becomes exact for the ground-state properties of the Holstein model. To facilitate this limit it is easiest to reintroduce units and make the transformation of harmonic oscillator coordinates $x \to \sqrt{2M\omega}x$, $p \to \sqrt{2/M\omega}p$. In real-space, the phonon terms in the Hamiltonian then become
\begin{equation}
    \Ham_{e-\mathrm{ph}}+\Ham_{\rm ph} = g\sqrt{2M\omega} \sum_i x_i \rho_i + \sum_i \frac{p_i^2}{2M} + \frac 12 M\omega^2 x_i^2.
\end{equation}
We take the limit $M\to\infty$, keeping the quantities $\alpha \equiv g \sqrt{2M\omega}$ and $k\equiv M\omega^2$ fixed, so that 
\begin{equation}
    \Ham_{e-\mathrm{ph}}+\Ham_{\rm ph} \to \alpha \sum_i x_i\rho_i + \sum_i \frac 12 k x_i^2.
\end{equation}
Note the dimensionless coupling is $\lambda = \alpha^2/2k$.

The ground-state configuration of the system is obtained by minimizing the energy functional
\begin{equation}
    E_0[\{x\}] = \varepsilon_0[\{x\}] + \sum_i \frac 12 k x_i^2
\end{equation}
with respect to the phonon coordinates $\{x\}$, where $\varepsilon_0$ is the ground-state energy of the electron part of the Hamiltonian in phonon configuration $\{x\}$. Minimizing $E_0$ yields the self-consistency condition  
\begin{equation}
    x_i = -\alpha\langle \rho_i \rangle/k,
    \label{eq:gen_self_consis}
\end{equation}
where we take the ground-state expectation value on the RHS.

Specializing to $\mathbf Q = (\pi,\pi)$ order we parametrize the phonon configuration as $x_i = x_0 + (-1)^{i_x + i_y}\delta x$, allowing also for a uniform shift. In this case, the self-consistency condition \eqref{eq:gen_self_consis} reduces to two equations for the $\qbf = 0$ and ${\mathbf Q}$ components:
\begin{equation}
    x_0 = -\alpha \rho/k, \quad \delta x = -\alpha \langle \rho_\mathbf Q \rangle/k.
    \label{eq:pipi_self_consis}
\end{equation}
Further specializing to the band-structure considered in the main text (nearest-neighbor hopping at half-filling) the first of these equations becomes $x_0 = -\alpha/k$ and half-filling corresponds to $\mu = -\alpha^2/k$. Putting everything into the Hamiltonian we obtain
\begin{equation}
    \Ham \!=\! \sum_{\kbf\sigma}\varepsilon_\kbf c^\dag_{\kbf\sigma}c_{\kbf\sigma}\! +\! \Delta\! \sum_{\kbf\sigma}\!c^\dag_{\kbf\! +\! \mathbf Q\sigma} c_{\kbf\sigma} \! +\! \frac{\alpha^2}{2k} N \!+\! \frac{k}{2\alpha^2}\Delta^2 N,
\end{equation}
where $\Delta \equiv \alpha\delta x$ and $N$ is the total number of lattice sites. Diagonalizing the electronic part of the Hamiltonian gives  
\begin{equation}
    H_e = \sideset{}{'}\sum_{\kbf\sigma} E_\kbf (\gamma^\dag_{\kbf \sigma +} \gamma_{\kbf \sigma +} - \gamma^\dag_{\kbf \sigma -} \gamma_{\kbf \sigma -}),
\end{equation}
where $E_\kbf = \sqrt{\varepsilon_\kbf^2 + \Delta^2}$, operators $\gamma_{\kbf\sigma\pm}$ are linear combinations of $c_{\kbf\sigma}$ and $c_{\kbf+\mathbf Q\sigma},$ and the prime indicates a sum over the reduced Brillouin zone defined by $|k_x| + |k_y| \leq \pi$. The $\qbf = \mathbf Q$ component of the density $\langle \rho_\mathbf Q \rangle$ can then be written
\begin{align}
   \frac 1N\!\sum_{\kbf \sigma}\langle c^\dag_{\mathbf k + \mathbf Q \sigma} c_{\mathbf k \sigma}\rangle= -\frac 1N\!\sideset{}{'}\sum_{\kbf \sigma}\! \frac{\Delta}{E_\kbf} \tanh\left(\frac{\beta E_\kbf}{2}\right).
\end{align}
Plugging this into \eqref{eq:pipi_self_consis} and seeking solutions with $\Delta \neq 0$ we obtain the ``gap equation"
\begin{equation}
    2\lambda \frac 1N \sideset{}{'}\sum_{\kbf \sigma} \frac{\tanh(\beta E_\kbf/2)}{E_\kbf} = 1.
    \label{eq:gap_eq}
\end{equation}
The result for $\Delta$ as a function of $\lambda$ on a 4$\times$4 lattice is shown in Fig.~\ref{fig:del_vs_lam}. Note we also introduce a small non-zero temperature to smooth out the singular Fermi functions at $T=0$.

The structure factor $N(\pi,\pi)$ is also readily obtained in the adiabatic limit as
\begin{equation}
    N(\pi, \pi) = N\langle \rho_\mathbf Q \rangle^2 + \frac 1N \sideset{}{'}\sum_{\kbf \sigma} \frac{1}{E_\kbf^2} \frac{\Delta^2 + \varepsilon_\kbf^2 \cosh(\beta E_\kbf)}{1 + \cosh(\beta E_\kbf)}.
\end{equation}
This is the formula used for $N(\pi,\pi)$ in Fig.~\ref{fig:holstein} of the main text.

\setcounter{figure}{0}
\renewcommand\thefigure{C\arabic{figure}}
\section{Convergence for Small Phonon Frequencies}\label{app:smallPhonon}
\begin{figure}[!th]
\begin{center}
\includegraphics[width=\columnwidth]{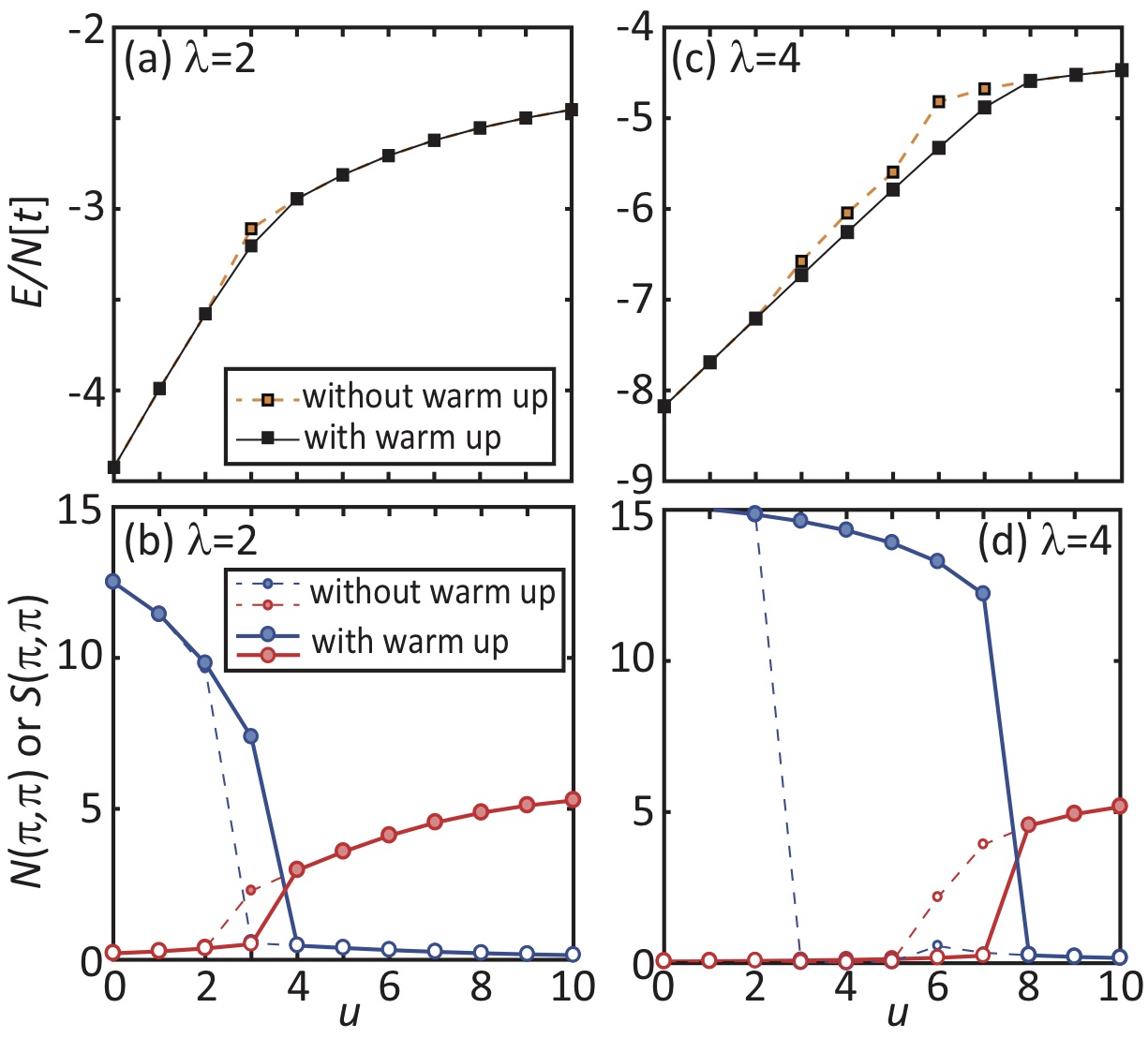}
\caption{\label{fig:phaseRelaxation} (a) The average energy of ground state calculated for various $u$ values and fixed $\lambda=2$. The orange (black) squares with dashed (solid) lines denote the final results without (with) a warm-up process. (b) The calculated spin (red) and charge (blue) structure factor for various $u$ values and $\lambda=2$, with (solid) and without (dashed) the warm-up process. (c,d) The same as (a,b) but for $\lambda=4$ instead.
}
\end{center}
\end{figure}
In this appendix, we present some detailed results about the small-frequency $\omega=t$ system and discuss the convergence issue in small-frequency systems.
Complementary to the cuts along the $\lambda$-axis, here in Fig.~\ref{fig:phaseRelaxation} we present two cuts along the $u$-axis with $\lambda=2$ and 4, respectively. Compared to the $\omega=5$ results in Fig.~\ref{fig:phaseDiag}, the small-frequency system exhibit a steeper transition near the phase boundary, due to the adiabatic reasoning mentioned in the main text.

It is worth to mention that the convergence is much harder for $\omega=t$ compared to larger frequencies. It typically takes 100-200 iterations even without reaching the close proximity of the phase boundary, while the $\omega=5t$ systems converge within 30 iterations. This is because the retardation of phonons drives the system away from an effective electronic model. The electron and phonon states have to exchange information many times to adjust to the optimal configuration. Near the phase boundary, the convergence can even be trapped by some local minima within the numerical accuracy $10^{-6}$, as shown in Fig.~\ref{fig:phaseRelaxation}. Due to the suppression of quantum fluctuations, the local minima barrier becomes steeper. To overcome this issue, we add ``warm-up'' iterations for larger $\omega$ but with the same $\lambda$ [\emph{i.e.},~using $g^\prime = \sqrt{\alpha} g$ and $\omega^\prime = \alpha \omega$ where $\alpha$ is a scaling factor much larger than 1]. These iterations are relatively faster and give raw approximations for the ground-state configurations at small frequencies, avoiding possible local minima. Figure~\ref{fig:phaseRelaxation} shows the results for the ground-state energy and structure factors obtained using and without using ``warm-up'' iterations. For systems near a phase transition, inappropriate treatment of the convergence may lead to a completely incorrect phase near the transition, though close in energy. The results in Fig.~\ref{fig:phaseVariousWU} were obtained by asymptotically tuning the scaling factor from 16, 8, 4, 2 to 1.

\begin{figure}[!th]
\begin{center}
\includegraphics[width=\columnwidth]{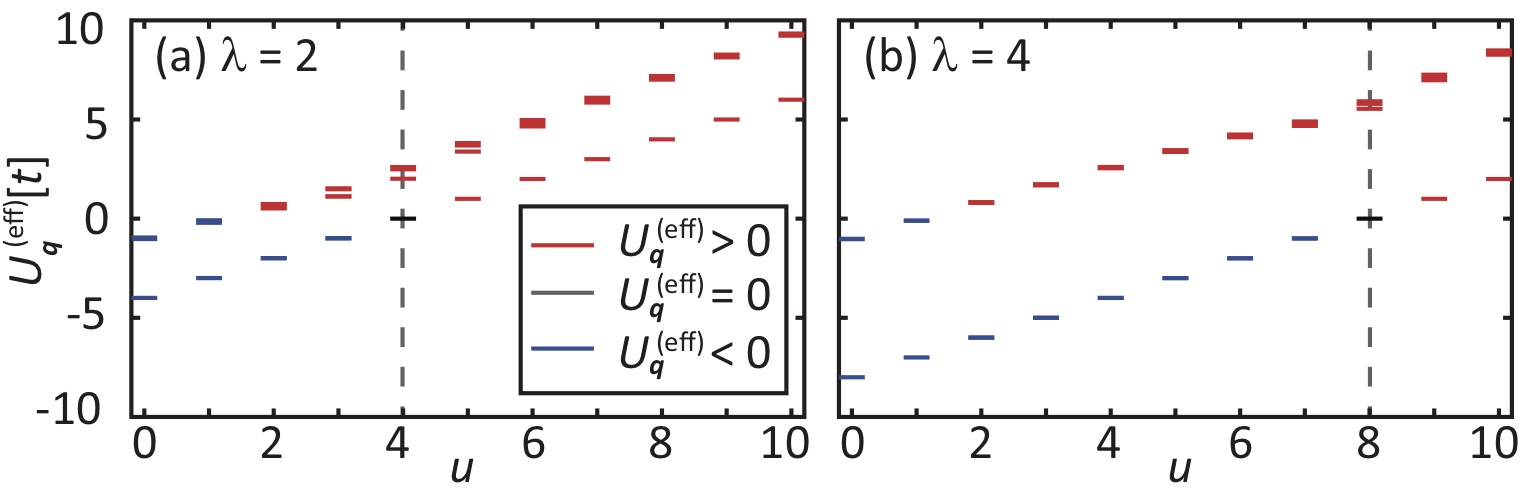}
\caption{\label{fig:vqDistrW1} Distribution of the effective interaction $U_\qbf^{\rm(eff)}$ for various $u$ values, with fixed (a) $\lambda=2$ and (b) $\lambda=4$. The phonon frequency is fixed as $\omega=1t$. The red bars represent $U_\qbf^{\rm(eff)}>0$, blue bars represent $U_\qbf^{\rm(eff)}<0$, and black bars represent $U_\qbf^{\rm(eff)}=0$. The dashed line indicates the anti-adiabatic limit $u=2\lambda$.
}
\end{center}
\end{figure}

In addition to the ground-state energy and structure factors, we also present the effective interaction $U_\qbf^{\rm (eff)}$ in Fig.~\ref{fig:vqDistrW1}. The interaction is more dispersive compared to large $\omega$'s, indicating the effective interactions mediated by the phonon become longer-range in the adiabatic limit. However, as mentioned in the main text, the suppression of the quantum fluctuations occurs exponentially; therefore, these interactions become semi-classical and lead to a sharp mean-field-like transition near the phase boundary.

\setcounter{figure}{0}
\renewcommand\thefigure{D\arabic{figure}}
\section{Comparison with NGS-GS Method}

\begin{figure}[!th]
\begin{center}
\includegraphics[width=\columnwidth]{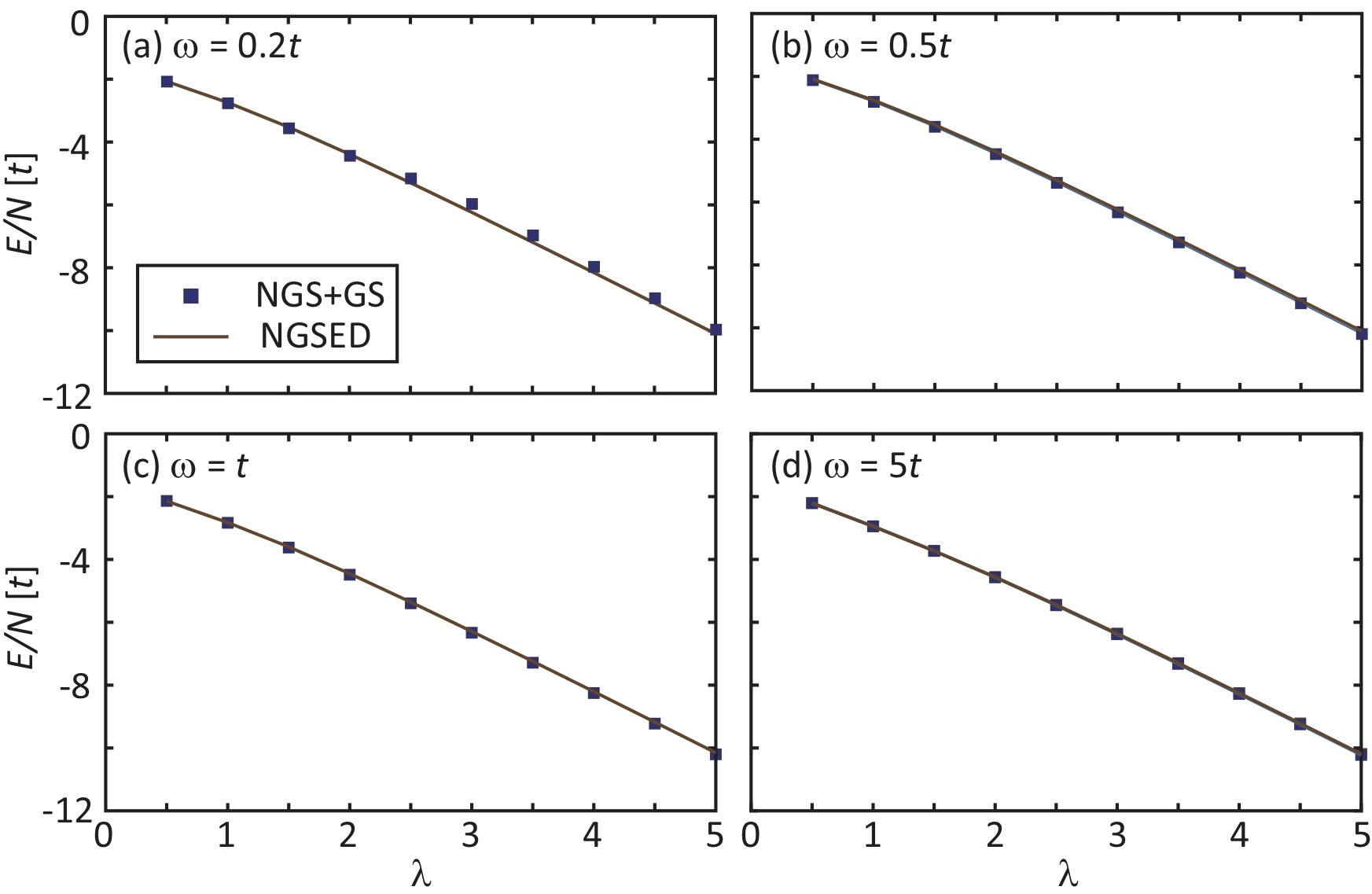}
\caption{\label{fig:NGScomp} Comparison of the ground-state energies obtained by NGSED calculations [solid lines, same as Fig.\ref{fig:holstein}(b)] and the g NGS + Gaussian wavefunction ansatz (blue squares), for phonon energies $\omega=0.2t$, 0.5$t$, $t$ and $5t$.
}
\end{center}
\end{figure}
In the limit of the Holstein model, previous studies have shown that the non-Gaussian transformation well describes the ground-state properties\,\cite{ohgoe2017competition, karakuzu2017superconductivity, shi2018variational}. To make a specific comparison, we present the calculation based on a pure variational ansatz
\begin{eqnarray}
    \big|\Psi(t)\big\rangle  = \Upl(t) |\psi_{\rm ph}^{\rm GS}\rangle \otimes|\psi_{\rm e}^{\rm GS}\rangle.
\end{eqnarray}
Here, the Gaussian phonon wavefunction $|\psi_{\rm ph}^{\rm GS}\rangle$ and the non-Gaussian transformation $\Upl$ are the same as the definition in Eqs.~\eqref{eq:wvfuncansatzPh} and \eqref{eq:nonGSansatz}. In contrast to the full many-body wavefunction $|\psi_{\rm e}\rangle$, the electronic part is also replaced by a Gaussian wavefunction
\begin{eqnarray}\label{eq:wvfuncansatzEle}
    |\psi_{e}^{\rm GS}\rangle =  e^{i \sum_{\ibf\jbf} \eta_{\ibf} c_{\ibf\uparrow}^\dagger c_{\jbf\downarrow}^\dagger} |0\rangle
\end{eqnarray}
The ground-state energies calculated using this non-Gaussian + Gaussian ansatz are summarized in Fig.~\ref{fig:NGScomp}. For most frequencies and coupling strengths, this ansatz is consistent with the results of NGSED, indicating that the electronic state indeed forms CDW orders in these cases. Only on the small-$\omega$ and large-$\lambda$ limits, the NGS+GS ansatz starts to deviate (slightly) from the NGSED. This can be attributed to the fact that the dressing factor $\lambda_\qbf$, in this case, becomes huge and causes stronger fluctuations.

\setcounter{figure}{0}
\renewcommand\thefigure{E\arabic{figure}}

\section{Details about Superconductivity}\label{app:superconductivity}
In this appendix, we discuss some details about the superconductivity in the 2D Hubbard-Holstein model. This includes the coherence of the Cooper pairs and different pairing symmetries.
\subsection{Coherence of Cooper Pairs}
For superconductivity, one important quantity is the coherence of the cooper pairs. To test this property, we also calculate the pairing correlation in the FFLO form. We extend the definition of the pairing operators to allow spatial modulation
\begin{eqnarray}
	 \Delta^{(s)}_\kbf &=& \sum_\kbf  c_{\kbf-\kprbf \downarrow} c_{\kprbf\uparrow}= \sum_\ibf  c_{\ibf\downarrow} c_{\ibf\uparrow}e^{-i\kbf\cdot\rbf_\ibf}
\end{eqnarray}
Then the pairing correlation determines the spatial coherence of Cooper pairs
\begin{eqnarray}
	 P^{(s)}_\qbf  &=& \sum_{\ibf,\jbf} \langle c_{\ibf\uparrow}^\dagger c_{\ibf\downarrow}^\dagger  c_{\jbf\downarrow} c_{\jbf\uparrow}\rangle e^{-i\qbf\cdot(\rbf_\jbf-\rbf_\ibf)}\nonumber\\
	 &=& \frac1{N}\sum_{\kbf^\prime} \langle\psi_e| \sum_{\kbf_2} c_{\kbf_2\downarrow}^\dagger  c_{\kprbf-\kbf_2\uparrow}^\dagger \sum_{\kbf_1} c_{\kprbf-\kbf_1\uparrow} c_{\kbf_1,\downarrow}|\psi_e\rangle\nonumber\\
	 && \sum_{\rbf}e^{-i(\kprbf-\qbf)\cdot \rbf}
	 e^{-\sum_\qbf\frac{4|\lambda_\qbf|^2}{N} e_2^T \Gamma_\qbf e_2 (1-e^{i\qbf\cdot\rbf})}
\end{eqnarray}
Note that $P^{(s)}_{\qbf=0}\equiv P_s$ in the BCS form.
The Fourier transform of $P^{(s)}_\qbf$ gives the real-space correlation of Cooper pairs at sites $\ibf$ and $\jbf$.

\begin{figure}[!th]
\begin{center}
\includegraphics[width=\columnwidth]{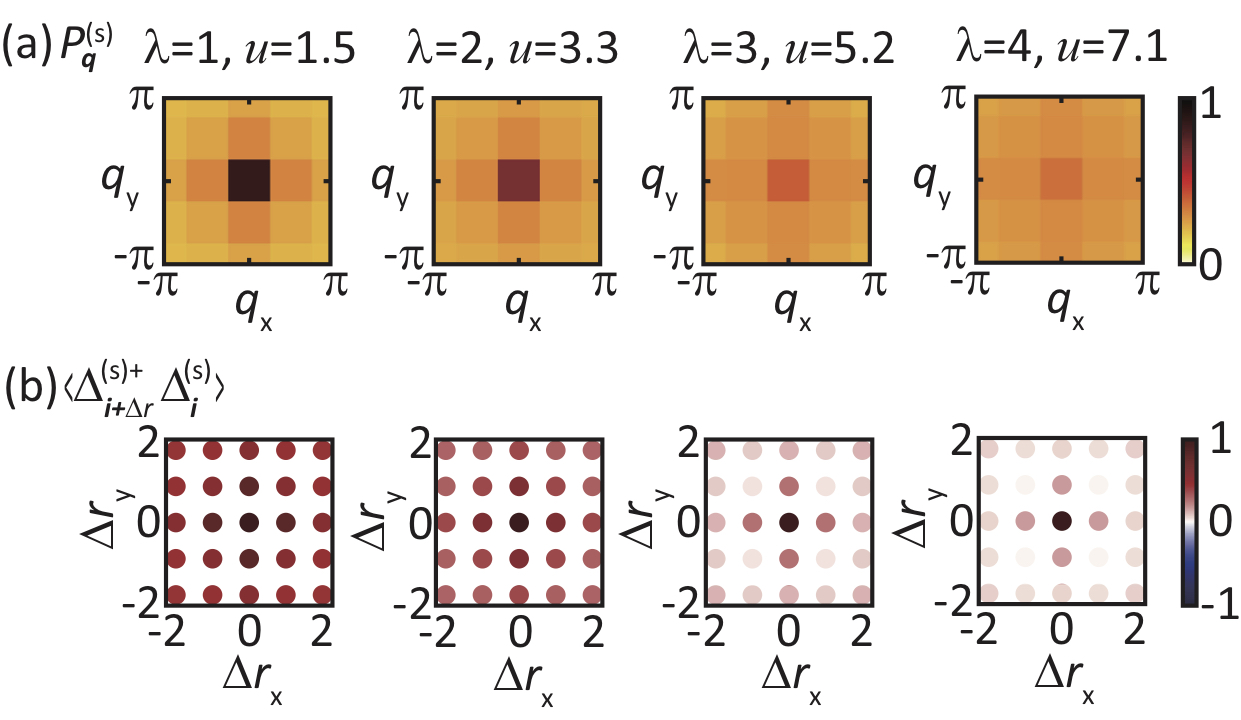}
\caption{\label{fig:pairCorr} (a) The FFLO pairing correlation calculated in the intermediate regime shown in Fig.~\ref{fig:DeltaSPairing}, \emph{e.g.} $\lambda=1$, $u=1.5$, $\lambda=2$, $u=3.3$, $\lambda=3$, $u=5.2$, and $\lambda=4$, $u=7.1$ respectively. (b) The real-space correlations of Cooper pairs as a function of distance, calculated for the same sets of parameters as (a).
}
\end{center}
\end{figure}
{\color{blue}
\begin{figure*}[!th]
\begin{center}
\includegraphics[width=15cm]{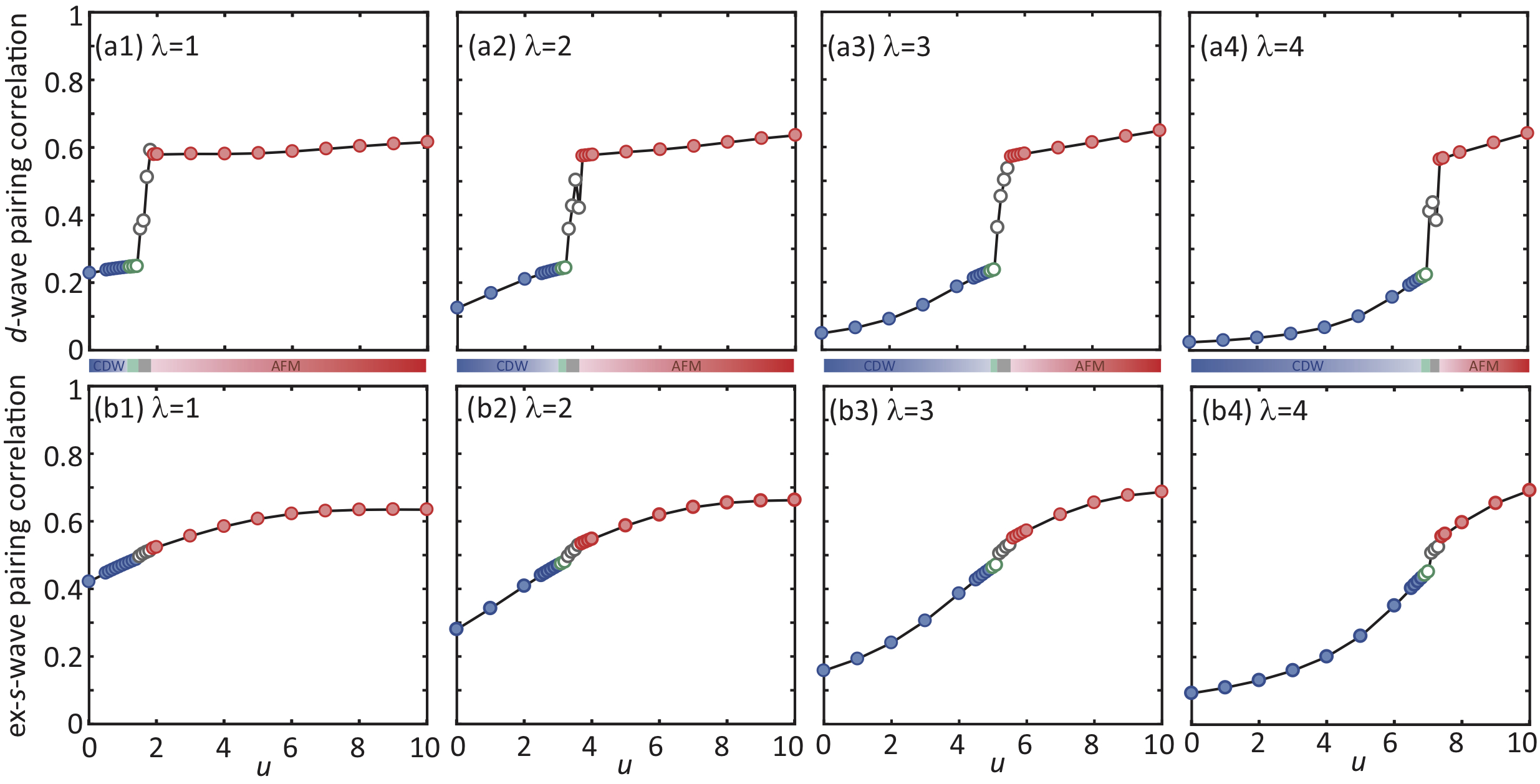}
\caption{\label{fig:pairingSymmetry} (a1-a4) $d$-wave and (b1-b4) extended-$s$-wave pairing correlations for $\lambda=1$, 2, 3, and 4, respectively. The color code follows the same convention of Figs.~\ref{fig:DeltaSPairing}(a-d).
}
\end{center}
\end{figure*}}

Figure \ref{fig:pairCorr} shows the FFLO pairing correlation and the real-space correlation for the intermediate phases for $\lambda=1$, 2, and 3. With the increase of the interaction parameters, the $P^{(s)}_{\qbf=0}$ decreases as explained in Fig.~\ref{fig:phaseDiag}. At the same time, the distribution of $P^{(s)}_{\qbf}$ in momentum space spreads out, reflecting that Cooper pairs are less coherent. This trend can also be reflected from the real-space correlation. As shown in Fig.~\ref{fig:pairCorr}(b), the Cooper pair correlation becomes shorter-range with the increase of interactions.

\subsection{Different Pairing Symmetries}

To test the possibility of other pairing symmetries, we also calculate the $d$-wave and extended-$s$-wave ($s^*$) pair correlation functions, defined as
\begin{eqnarray}
P_{d/s^*}\!  =  \sum_{\ibf,\jbf}\sum_{\alpha,\beta\atop\delta_\alpha,\delta_\beta} \langle c_{\ibf+\delta_\alpha\uparrow}^\dagger c_{\ibf\downarrow}^\dagger  c_{\jbf\downarrow} c_{\jbf+\delta_\beta\uparrow}\rangle f^{(d/s^*)}_{\alpha\beta}
\end{eqnarray}
Here, $\alpha$ and $\beta$ take $x$ or $y$ directions; the neighboring distance $\delta_{x}=\pm \hat{x}$ and $\delta_{y}=\pm \hat{y}$. The $d$-wave and extended-$s$-wave shape functions are 
\begin{eqnarray}
f^{(d/s^*)}_{\alpha\beta} = \left\{\begin{array}{ll} \frac14 &,\textrm{for}\ \alpha=\beta \\ \mp \frac14 &,\textrm{for}\ \alpha\neq\beta\end{array}\right.
\end{eqnarray}

Due to the non-local dressing with phonons, the $d$- and extended-$s$-wave pair correlation functions become more complicated upon making the unitary transformation:
\begin{widetext}
\begin{eqnarray}
P_{d/s^*}\!  &=&  \frac1N\sum_{\kprbf} \sum_{\kbf_1,\kbf_2}\xi(\kprbf,\kbf_1,\kbf_2) \left[\zeta_{x}(\kprbf)\cos\left(\frac{\kpr_x}{2} - k_{1x}\right) \cos\left(\frac{\kpr_x}{2} - k_{2x}\right)+ \zeta_{y}(\kprbf)\cos\left(\frac{\kpr_y}{2} - k_{1y}\right) \cos\left(\frac{\kpr_y}{2} - k_{2y}\right)\right.\nonumber\\
	 &&\left.\mp\zeta_{xy}(\kprbf)\cos\left(\frac{\kpr_x}{2}-k_{2x}\right) \cos\left(\frac{\kpr_y}{2}-k_{1y}\right)\mp\zeta_{xy}(\kprbf)\cos\left(\frac{\kpr_y}{2}-k_{2y}\right) \cos\left(\frac{\kpr_x}{2}-k_{1x}\right) \right]
\end{eqnarray}

In the last step, the bare electronic pairing correlation is 
\begin{eqnarray}
	 \xi(\kprbf,\kbf_1,\kbf_2)=\langle\psi_e|  c_{\kbf_2\downarrow}^\dagger c_{\kprbf-\kbf_2\uparrow}^\dagger   c_{\kprbf-\kbf_1\uparrow} c_{\kbf_1\downarrow} |\psi_e\rangle
\end{eqnarray}
and dressing factor 
\begin{eqnarray}
	 \zeta_{x}(\kprbf) = \sum_{\rbf} e^{-i\kprbf\cdot\rbf}e^{-\sum_\qbf\left[(1-\cos q_x)-e^{i\qbf\cdot\rbf}(1+\cos q_x)\right]  \frac{2|\lambda_\qbf|^2}{N}e_2^T \Gamma_\qbf e_2}\nonumber\\
	  \zeta_{y}(\kprbf) = \sum_{\rbf} e^{-i\kprbf\cdot\rbf}e^{-\sum_\qbf\left[(1-\cos q_y)-e^{i\qbf\cdot\rbf}(1+\cos q_y)\right]  \frac{2|\lambda_\qbf|^2}{N}e_2^T \Gamma_\qbf e_2}\nonumber\\
	  \zeta_{xy}(\kprbf) = \sum_{\bar{\rbf}}e^{-i\kprbf\cdot\bar{\rbf}}e^{-\sum_\qbf\left[(2-\cos q_x-\cos q_y)-4e^{i\qbf\cdot\bar{\rbf}}\cos\frac{q_x}{2}\cos\frac{q_y}{2}\right]  \frac{|\lambda_\qbf|^2}{N}e_2^T \Gamma_\qbf e_2} 
\end{eqnarray}
\end{widetext}
Here, the $\bar{\rbf}$ in the last summation denotes the half-unit-cell-shifted coordinates $\bar{\rbf} = \rbf + \hat{x}/2 - \hat{y}/2$.

As shown in Fig.~\ref{fig:pairingSymmetry}, both correlations increase with $u$ since the on-site Coulomb interactions favor non-local pairs. In contrast to the $s$-wave superconductivity, none of these pair correlators display a sharp peak in the intermediate regime that is larger than the correlations in the AFM phase. Therefore, we believe the dominant pairing symmetry is $s$-wave in the intermediate regime.

\end{document}